\documentclass[aps,preprint,pre,superscriptaddress,letterpaper,amsmath,amssymb]{revtex4-1}

\usepackage[T1]{fontenc} 
\usepackage{graphicx}
\usepackage{graphics}
\usepackage[mathscr]{eucal}
\usepackage{amssymb}
\usepackage{amsmath}
\usepackage{xspace}
\usepackage{listings}
\usepackage{color}
\usepackage{bm}         
\usepackage{array}
\usepackage{multirow}   
\usepackage{booktabs}   
\usepackage{siunitx}
\usepackage{ marvosym } 
\usepackage{setspace}
\usepackage[caption=false]{subfig} 

\usepackage{verbatim}

\usepackage{subfig}

\usepackage{mleftright}
\usepackage[utf8]{inputenc}
\newcommand{\e}{\mathrm e}
\renewcommand{\d}{\mathrm d}
\renewcommand{\i}{\mathrm i}
\newcommand*{\citen}{}
\DeclareRobustCommand*{\citen}[1]{%
  \begingroup
    \romannumeral-`\x 
    \setcitestyle{numbers}%
    \cite{#1}%
  \endgroup
}

\begin{document}



\title{Subcompartmentalization of polyampholyte species in organelle-like condensates is promoted by charge pattern mismatch and strong excluded-volume interaction}

\date{March 8, 2021}

\author{Tanmoy~Pal}
\affiliation{Department of Biochemistry, University of Toronto, Toronto, Ontario M5S 1A8, Canada}

\author{Jonas~Wess\'en}
\affiliation{Department of Biochemistry, University of Toronto, Toronto, Ontario M5S 1A8, Canada}

\author{Suman~Das}
\affiliation{Department of Biochemistry, University of Toronto, Toronto, Ontario M5S 1A8, Canada}

\author{Hue~Sun~Chan}
\affiliation{Department of Biochemistry, University of Toronto, Toronto, Ontario M5S 1A8, Canada}
\affiliation{{\rm To whom correspondence should be addressed. Email:}
{\tt chan@arrhenius.med.utoronto.ca}}

\begin{abstract}
Polyampholyte field theory and explicit-chain molecular dynamics 
models of sequence-specific phase separation of a system 
with two intrinsically 
disordered protein (IDP) species indicate consistently that a 
substantial polymer excluded volume and a significant mismatch of 
the IDP sequence charge patterns can act in concert, but not in 
isolation, to demix the two IDP species upon condensation. This 
finding reveals an 
energetic-geometric interplay in 
a stochastic, ``fuzzy'' molecular recognition mechanism that may 
facilitate subcompartmentalization of membraneless organelles.

\end{abstract}


\maketitle


\section{Introduction}

Liquid-liquid phase separation 
(LLPS) \cite{Brangwynne2009,Rosen12,McKnight12,CellBiol,Nott2015} in
biomolecular condensates~\cite{BananiLeeHymanRosen2017} has garnered
intense interest in diverse areas of biomedicine, biophysics,
and polymer physics~\cite{ALBERTI2017R1097}. 
LLPS plays a central role in the assembly of droplet-like cellular 
compartments---coexisting with a more dilute milieu and sometimes 
referred to as membraneless 
organelles---that act as hubs for biochemical processes.
Examples include nucleoli, 
P-bodies, stress granules, and cajal bodies.
Serving critical organismal functions, their misregulation can cause
disease~\cite{MolliexTemirovLeeCoughlinKanagarajKimMittagTaylor,
LiChavaliPancsaChavaliBabu}.

Biomolecular LLPS often involves intrinsically disordered 
proteins (IDPs) and nucleic acids participating in multivalent
interactions \cite{cosb15,NatPhys}. 
Recent theories and computational studies have begun to shed light on how 
LLPSs of IDPs are governed by their amino acid sequences. These efforts 
include analytical theory \cite{LinPRL,SingPerry2017,Kings,alanamin2020},
explicit-chain
lattice \cite{FERICNucleolus,DasEisenLinChan,ChoiDarPappu2019}
and continuum molecular dynamics (MD) 
\cite{jeetainPLOS,DasAminLinChan,panag2020,koby2020,sumanPNAS2020,koby2021}
simulations, and
field-theoretic simulation (FTS) \cite{LinMcCartyRauchDelaneyKosikFredricksonSheaHan2019,McCartyDelaneyDanielsenFredricksonShea2019,FredricksonPNAS},
investigations of the relationship between LLPS propensity
and single/double-chain properties \cite{LinChan2017,Dignon2018,alanamin2020}
as well as crystals and filaments formation~\cite{Wallin2019},
and studies of the peculiar
temperature~\cite{rolandJACS2019,jeetainACS} and 
pressure \cite{rolandJACS2019,roland20}
dependence of biomolecular LLPS
as well as finite-size scaling in droplet formation~\cite{Anders2020}.
Reviews of the emerging theoretical perspectives are available in 
Refs.~\citen{LinKayChan,jeetainRev,Roland2019,choiRev2020,SingPerryRev2020}.

IDPs are enriched in charged and polar residues \cite{uversky2002} and
multivalent electrostatics is an important driving force---among
others \cite{Robert-Julie}---for LLPS.
One consistent finding from theory \cite{LinPRL}, 
chain simulation \cite{DasEisenLinChan,DasAminLinChan}~and
FTS \cite{McCartyDelaneyDanielsenFredricksonShea2019,FredricksonPNAS}
is that the LLPS propensity of a polyampholyte depends on its sequence
charge pattern, which may be quantified by an intuitive blockiness 
$\kappa$ measure \cite{DasPappu2013} or an analytic ``sequence 
charge decoration'' (SCD) parameter \cite{SawleGhosh2015} that correlates
with single-chain properties \cite{DasPappu2013,SawleGhosh2015,HuihuiGhosh2020}.

While simple laboratory systems may contain only one 
IDP type (species), many types of IDPs 
interact in the cell to compartmentalize into
a variety of condensates.
In some cases, LLPS-mediated organization of intracellular 
space goes a step further by subcompartmentalization \cite{ThiryLafontaine}. 
Well-known examples include the nucleolus comprising of at least three 
subcompartments enriched with distinct sets of 
proteins \cite{ncrna5040050,FERICNucleolus} and stress granules with a 
dense core surrounded by a liquid-like outer shell \cite{JAIN2016487}. These 
phenomena 
raise intriguing physics questions as to the nature
of the sequence-specific interactions that drive a subset of IDPs in a 
condensate to coalesce among themselves while excluding other types of IDPs.

Insights into formation of 
subcompartments \cite{FERICNucleolus,harmonNJP,HXZhou2021} 
and general principles 
of many-component phase behaviors \cite{Frenkel2017}
have been gained from models with energies assigned to 
favor or disfavor interactions between different solute components.
On a more fundamental level, a random phase approximation 
(RPA) \cite{delacruz,LinPRL} model of two polyampholytic 
IDP species suggested that sequence-specific molecular recognition can
arise from elementary electrostatic interactions in a stochastic, ``fuzzy'' 
manner, in that the IDP species in the LLPS condensed phase are 
predicted to demix when their sequence charge patterns are significantly 
different (large difference in their
SCD values), but tend to be miscible when 
their SCD values are similar~\cite{LinNewJPhys}. This trend is
also rationalized by a recent analysis of second
virial coefficients~\cite{alanamin2020}.


\begin{table*}[ht!]
\caption{Hamiltonians
used in this work; $\beta = 1/k_{\mathrm{B}} T$, where
$k_{\mathrm{B}}$ is Boltzmann's constant and $T$ is absolute
temperature.}\label{tab1}
\begin{ruledtabular}
\vskip 2mm

{\footnotesize{
\begin{tabular}{lccc}
{}      & {$\hat{H}_0$} & {$\hat{H}_1$} & {$\hat{H}_2$}\\ \hline
FTS: &
{\large $\frac {3}{2 b^2 \beta}$}
\hskip -1mm
{$
\sum\limits_{p,i,\alpha}
\vert\bm{r}_{p,i,\alpha+1} - \bm{r}_{p,i,\alpha}\vert^2 $}
&
{\large $\frac {v}{2\beta}$}
\hskip -1mm
{$
\int \d \bm{r}\int\d\bm{r}'
\hat{\rho}_{\mathrm{b}}(\bm{r})\delta (\bm{r}-\bm{r}')
\hat{\rho}_{\mathrm{b}}(\bm{r}')$}
&
\hskip -4mm
{\phantom{$\Biggl\vert$}}
{\large ${\frac{l_{\mathrm{B}}}{2\beta}}$}
\hskip -1mm
{$
\int \d \bm{r} \int \d \bm{r}'
$}
{\large
$
\frac{\hat{\rho}_{\mathrm{c}}(\bm{r})
\hat{\rho}_{\mathrm{c}}(\bm{r}') }{|\bm{r} - \bm{r}'|} $
}
\\
{MD:}
&
{\large $\frac {K_{\mathrm{b}}}{2}$}
\hskip -2mm
{$
\sum\limits_{p,i,\alpha}
(|\bm{r}_{p,i,\alpha+1} - \bm{r}_{p,i,\alpha}| - a_0 )^2 $}
&
\hskip 1.5mm
{\large $\frac {2\epsilon}{3}$}
\hskip -2.5mm
{$
\sum\limits_{\substack{p,i,\alpha, \\ \neq q,j,\gamma}}
$}
{$
\hskip -2mm
\Bigl[
\Bigl($
\hskip -1.5mm
{\large
$\frac{r_0}{|\bm{r}_{p,i,\alpha}-\bm{r}_{q,j,\gamma}|} $
}
\hskip -3.0mm $\Bigr)^{12}
 -
\Bigl($
\hskip -1.5mm
{\large
$\frac{r_0}{ |\bm{r}_{p,i,\alpha} - \bm{r}_{q,j,\gamma}|} $
}
\hskip -3.0mm $\Bigr)^{6}
\Bigr]$}
&
\hskip -3mm
{\large $\frac {l_{\mathrm{B}}}{2\beta}$}
\hskip -2mm
{$
\sum\limits_{\substack{p,i,\alpha, \\
\neq q,j,\gamma}}$}
\hskip -2mm
{\large $\frac{\sigma_{p,\alpha} \sigma_{q,\gamma}}{|\bm{r}_{p,i,\alpha}
- \bm{r}_{q,j,\gamma}|}$}\\
\end{tabular}
}}
\end{ruledtabular}
\vskip -2mm
\end{table*}


Aiming to better understand the physics of selective
compartmentalization in membraneless organelles,
a question that must be tackled is how sequence charge pattern 
and polymer excluded volume interplay in the mixing/demixing of 
condensed polyampholyte species. 
The question arises because excluded volume was not fully accounted 
for in RPA~\cite{LinNewJPhys} but
excluded volume is a known factor in 
LLPS~\cite{DasAminLinChan,McCartyDelaneyDanielsenFredricksonShea2019}
and other condensed-phase properties~\cite{adro2017,sorichetti2018}.
In the present work, we address this fundamental question
by using FTS and MD
to model polyampholytes with short-range excluded 
volume repulsion and long-range Coulomb interaction. 
By construction, FTS is more accurate than RPA in the field-theoretic
context if discretization and finite-volume errors 
can be neglected, whereas MD is more suitable for chemically realistic
interactions 
and its microscopic structural information is accessible.
As shown below, both models indicate that
while charge pattern mismatch is necessary for demixing of different 
polyampholyte species in the condensed phase, 
the degree of demixing is highly sensitive to excluded volume, 
underscoring that excluded volume is a critical
organizing principle not only for folded protein 
structures \cite{ChanDill90,Banavar00,eis2006} 
and disordered protein conformations
\cite{WallinTopo2005,Wallin2006,Song2015} but also for biomolecular 
condensates.

\section{Model and Rationale}

\begin{figure}[!ht]
\includegraphics{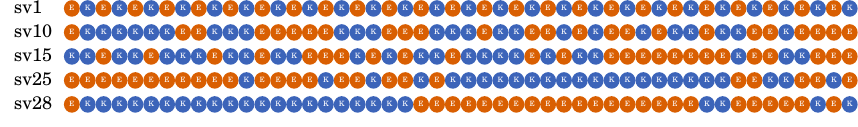}
\vskip -3mm
\caption{Polyampholytes studied in this work.
Blue/red beads of ``K''s (lysines)/``E''s (glutamic acids)
carry $\pm 1$ protonic charges. The sv labels 
are those of Ref.~\citen{DasPappu2013}.}
\label{fig:sequences}
\end{figure}

Here we study binary mixtures of two species of 
fully charged, overall neutral bead-spring polyampholytes $p,q$ differing only 
in their charge patterns, defined by the set of positions
$\bm{r}_{p,i,\alpha}$ ($\bm{r}_{q,i,\alpha}$) 
of bead $\alpha$ on chain $i$ of type 
$p$ ($q$) with electric charges $\sigma_{p,\alpha}$ ($\sigma_{q,\alpha}$) for 
all $i$. The sequences considered (Fig.~\ref{fig:sequences})
are representative of the set of 50mer ``sv sequences'',
used extensively for modeling \cite{SawleGhosh2015,LinChan2017,LinNewJPhys,McCartyDelaneyDanielsenFredricksonShea2019}, that are listed in ascending 
$\kappa$ values from the least blocky, strictly alternating sv1
to the diblock sequence sv30 \cite{DasPappu2013}. 
As a first step in studying pertinent general principles,
the simple, coarse-grained FTS and MD Hamiltonians
$\hat{H} = \hat{H}_0 + \hat{H}_1 + \hat{H}_2$
in Table~\ref{tab1} are adopted without consideration of
structural details and variations such as salt and pH dependence.
While all of our model sequences have net zero charge and thus counterions
are not needed to maintain overall neutrality of the system,
experiments show that formation of biomolecular condensates is affected by
salt and pH~\cite{ALBERTI2017R1097,jacobPNAS2017}. Recently, some of these
effects are rationalized by an improved RPA formulation with renormalized 
Kuhn lengths for the LLPS of a single polyampholyte species~\cite{Kings}.
The study of these effects should be extended to multiple IDP species
in future efforts.

Following standard prescription, we have expressed $\hat{H}_1$ and 
$\hat{H}_2$ in terms of {\hbox{$\hat{\rho}_{\mathrm{b}}(\bm{r}) =  
\sum_p \hat{\rho}_{\mathrm{b},p}(\bm{r})$}},
$\hat{\rho}_{\mathrm{c}}(\bm{r}) =  \sum_p \hat{\rho}_{\mathrm{c}, p}(\bm{r}) $
where $\hat{\rho}_{\mathrm{b},p}$ and $\hat{\rho}_{\mathrm{c},p}$ are,
respectively, the microscopic bead (matter) and charge densities
of polymer type $p$. 
The individual beads are modelled as normalized Gaussian distributions
$\Gamma(\bm{r})=\exp(-\bm{r}^2/2a^2)/(2\pi a^2)^{3/2}$
centered at positions $\bm{r}_{p,i,\alpha}$
\cite{RigglemanRajeevFredrickson2012,Wang2010} such that $
\hat{\rho}_{\mathrm{b},p}(\bm{r}) = \sum_{i,\alpha} \,\,
\Gamma(\bm{r}-\bm{r}_{p,i,\alpha}) $, 
$ \hat{\rho}_{\mathrm{c},p}(\bm{r}) =
\sum_{i,\alpha}\sigma_{p,\alpha} \Gamma(\bm{r}-\bm{r}_{p,i,\alpha}) $.
The chain connectivity term
$\hat{H}_0$ takes the usual Gaussian form with Kuhn length $b$ 
for FTS and the harmonic form with force constant $K_{\mathrm{b}}$
for MD (thus $b$ corresponds to $a_0$); the excluded-volume 
term $\hat{H}_1$ entails a $\delta$-function with strength $v$
for FTS \cite{Edwards1965,McCartyDelaneyDanielsenFredricksonShea2019} 
and a Lennard-Jones (LJ) potential with well depth $\epsilon/3$ for MD
\cite{DasAminLinChan}; whereas electrostatics is provided by 
$\hat{H}_2$ with Bjerrum length 
$l_{\mathrm{B}}=e^2/4\pi\epsilon_0\epsilon_{\rm r}k_{\rm B}T$,
where $e$ is electronic charge, $\epsilon_0$ and $\epsilon_{\rm r}$ are,
respectively, vacuum and relative permitttivity (larger $l_{\mathrm{B}}$
corresponds to stronger electrostatic interactions because of a smaller
$\epsilon_{\rm r}$ and/or lower $T$).


The phase behavior and the mixing/demixing of fully charged, overall-neutral
polyampholytic sv sequences~\cite{DasPappu2013} in the condensed phase 
are used here as an idealized system to investigate the electrostatic aspects 
of the driving forces for these phenomena. For real systems of biological 
or synthetically designed IDPs, other favorable interactions~\cite{NatPhys},
including non-ionic and 
hydrophobic~\cite{rolandJACS2019,jeetainACS,knowles2021} and
$\pi$-related~\cite{Robert-Julie,sumanPNAS2020,Songetal2013} effects, can 
afford additional contributions to the stability of the condensed phase.
Thus, the behavior of a model system simulated here at a given model
temperature (a given $l_{\mathrm{B}}$) may correspond to that of a 
system of IDPs with similar electrostatic but additional favorable 
physical interactions
at a higher experimental temperature (shorter $l_{\mathrm{B}}$).
Bearing this in mind, we choose $l_{\mathrm{B}}=5b$ to obtain many of the 
FTS results presented below in order to ensure that the model sequence with the 
lowest LLPS propensity, namely sv1, would phase separate, because we
are interested primarily in the mixing/demixing of different IDP species
in the condensed phase. In other words, $l_{\mathrm{B}}=5b$ is lower
than the upper critical solution temperatures (UCST) of all the sequences
we consider. If we take $T=298.15$ K as room temperature
and $b$ $=$ C$_\alpha$--C$_\alpha$ 
virtual bond length $=$ $3.8$~\AA, $l_{\mathrm{B}}=5b$ corresponds to
a relative permittivity $\epsilon_{\rm r}\approx 30$ for the
solvent plus IDP environment. While $l_{\mathrm{B}}=5b$ is larger than
$l_{\mathrm{B}}\approx 1.8b\approx 7$\AA~if the 
$\epsilon_{\rm r}\approx 80$ for bulk water is assumed,
it is instructive to note that the dielectric environment of the 
IDP condensed phase likely entails a smaller effective $\epsilon_{\rm r}$ than
that of bulk water~\cite{sumanPNAS2020,Wessen_etal2021}, and that
uniform relative permittivities with
$\epsilon_{\rm r}$ values of $\approx 30$--$60$
have been used recently to match theoretical predictions with
experimental LLPS data~\cite{Nott2015,LinPRL,LinJML}.


\section{Field theoretic simulations (FTS)}

The basic strategy of field-based approaches is to trade the explicit bead positions $\lbrace \bm{r}_{p,i,\alpha} \rbrace $ in favor of a set of interacting fields as the microscopic degrees of freedom (the mathematical procedure to achieve this is outlined below). The resulting statistical field theory contains the same thermodynamic information, and therefore thermal averages over any function of bead positions $\hat{\mathcal{O}}(\lbrace \bm{r}_{p,i,\alpha} \rbrace)$ can in principle always be computed as field averages of some field operator $\tilde{\mathcal{O}}$, although finding the corresponding $\tilde{\mathcal{O}}$ for a given $\hat{\mathcal{O}}$ is far from trivial if $\hat{\mathcal{O}}$ has a complicated dependence of $\lbrace \bm{r}_{p,i,\alpha} \rbrace$. To distinguish these two types of averages in this section, we let $\langle
\dots \rangle_{\mathrm{P}} $ and $\langle \dots \rangle_{\mathrm{F}} $ denote,
respectively, averages over bead centers (i.e., in the ``particle picture'') 
and averages over field configurations (i.e., in the ``field picture'').

Because explicit bead positions are not readily available in the field picture, spatial information about the chains has to be gleaned from 
functionals of $\lbrace \hat{\rho}_{\mathrm{b},p} \rbrace$ that have well-defined corresponding field operators. A set of such quantities are the pair-distribution functions (PDFs), 
\begin{equation}
\label{eq:cross_correlation_particle}
G_{p,q}(|\bm{r}-\bm{r}'|) = \langle \hat{\rho}_{\mathrm{b},p}(\bm{r}) \hat{\rho}_{\mathrm{b},q}(\bm{r}') \rangle_{\mathrm{P}} \, , 
\end{equation}
between various $p,q$ bead types. That $G_{p,q}$ depends only on
$|\bm{r}-\bm{r}'|$ follows from translational and rotational invariance. Both inter- 
($p\neq q$) and intra ($p=q$) species PDFs are needed
to characterize structural organization of different species.
For instance, an intra species $G_{p,p}(r)$ peaking at small $r$ and
decaying to $0$ at large $r$ implies a relatively dense region, i.e., a droplet,
of $p$; and demixing of two species $p$ and $q$ is signalled by
$G_{p,p}(r)$ and $G_{q,q}(r)$ dominating over $G_{p,q}(r)$ at small $r$. As noted in the Appendix, more accurate spatial 
information is provided by $G_{p,q}$s than by perturbative 
second virial coefficients~\cite{alanamin2020,Pathria,lenhoff1998}.

Following standard methods (see e.g.~\cite{Fredrickson2006} for detailed formulation), we now show how to derive the field theory of our model, and then how PDFs can be computed in the field picture. We begin by considering the canonical 
partition function expressed as integrals over the positions of bead centers,
$\bm{r}_{p,i,\alpha}$, in the particle picture, 
with an added source field $J_p(\bm{r})$ for each bead type density as
is commonly practiced 
in field theory to facilitate subsequent calculation of averages of
functionals of $\hat{\rho}$:
\begin{equation} \label{eq:particle_can_ensemble}
Z[ \lbrace J_p \rbrace ] = \left( \prod_{p,i,\alpha}  \int \d
\bm{r}_{p,i,\alpha} \right) \e^{-\beta \hat{H}_0 - \beta \hat{H}_1 - \beta
\hat{H}_2 + \int \d \bm{r} \sum_{p} \hat{\rho}_{\mathrm{b},p}(\bm{r})
J_p(\bm{r}) } \, .  
\end{equation}
The FTS interaction strengths are controlled~by 
$v$ and $l_{\mathrm{B}}$ (Table~\ref{tab1}). 
To minimize notational clutter, overall multiplicative constant factors in $Z$ 
that are immaterial to the quantities computed in this work are not
included in the mathematical expressions in the present derivation. 
Using Eq.~\eqref{eq:particle_can_ensemble}, averages of products of
bead densities can formally be computed using functional derivatives of $Z$
with respect to the source fields $J_p$, then followed by setting $J_p=0$ 
for all $p$. In
particular, 
\begin{equation} \label{eq:PDF_from_source}
G_{p,q}(|\bm{r}-\bm{r}'|)  = 
\lim_{J_p,J_q \rightarrow
0} \hskip 1mm \frac{1}{Z} \frac{\delta}{\delta J_p(\bm{r})} \frac{\delta}{\delta
J_q(\bm{r}')} Z \, .  
\end{equation} 
To derive the field theory, we first multiply the right hand 
side of Eq.~\eqref{eq:particle_can_ensemble} (from the left) by unity (`1') in the form of
\begin{equation}
1 = \int \mathcal{D} \rho_{\mathrm{b}}(\bm{r}) \, \delta[ \rho_{\mathrm{b}}-\hat{\rho}_{\mathrm{b}}] \, \int \mathcal{D} \rho_{\mathrm{c}}(\bm{r}) \, \delta[\rho_{\mathrm{c}}-\hat{\rho}_{\mathrm{c}}] \, , 
\end{equation}
after which we can make the replacements $\hat{\rho}_{\mathrm{b,c}} \rightarrow
\rho_{\mathrm{b,c}}$ in $\beta \hat{H}_{1,2}$ because of the 
$\delta$-functionals. The $\delta$-functionals are then expressed in their equivalent
Fourier forms, 
\begin{equation}
\delta[\rho_{\mathrm{b}}-\hat{\rho}_{\mathrm{b}}] = \int \mathcal{D} w (\bm{r})
\,  \e^{\i \int \d \bm{r} w(\rho_{\mathrm{b}} - \hat{\rho}_{\mathrm{b}}) }
\; , \quad \quad \delta[\rho_{\mathrm{c}}-\hat{\rho}_{\mathrm{c}}] = \int
\mathcal{D}\Phi (\bm{r}) \, \e^{\i \int \d \bm{r} \Phi(\rho_{\mathrm{c}} -
\hat{\rho}_{\mathrm{c}}) } \, ,  
\end{equation}
where $\i^2=-1$,
to allow for the explicit functional integrals over 
the $\rho_{\mathrm{b}}(\bm{r})$ and $\rho_{\mathrm{c}}(\bm{r})$ 
variables introduced by the above `1' factor.
Up to a multiplicative constant, the result of those integrations is
the formula
\begin{equation} \label{eq:field_can_ensemble}
Z[\lbrace J_p \rbrace ] = \int \mathcal{D}w(\bm{r}) \int
\mathcal{D}\Phi(\bm{r}) \,\, \e^{-H[w,\Phi; \lbrace J_p \rbrace ]} \, ,
\end{equation}
where the field Hamiltonian is
\begin{equation} \label{eq:H}
H[w,\Phi; \lbrace J_p \rbrace] = - \sum_p n_p \ln Q_p[\i \breve{w} -
\breve{J}_p,\i \breve{\Phi} ] + \int \d\bm{r} \left( \frac{w^2}{2 v} +
\frac{(\bm{\nabla} \Phi)^2}{8 \pi l_{\mathrm{B}} } \right) \, ,
\end{equation}
and $\breve{w}(\bm{r}) \equiv \Gamma \star w(\bm{r}) \equiv \int \d \bm{r}' \Gamma(\bm{r}-\bm{r}') w(\bm{r}')$ (and similarly for $\breve{\Phi}$ and $\breve{J}_p$).
Here,  $Q_p[\i\breve{w},\i \breve{\Phi}]$ is the partition function of a single polymer of
type $p$, subject to external chemical and electrostatic potential fields $\i
\breve{w} $  and $\i \breve{\Phi}$, respectively, i.e.~
\begin{equation}
Q_p[\i \breve{w},\i \breve{\Phi} ] \equiv 
\frac {1}{V} \left( \frac {3}{2\pi b^2}\right)^{{3(N_p-1)}/{2} }
\left(
\prod_{\alpha=1}^{N_p} \int \d \bm{r}_\alpha \right) \exp\left[ - \frac{3}{2b^2}
\sum_{\alpha=1}^{N_p-1} \left(\bm{r}_{\alpha+1} - \bm{r}_{\alpha} \right)^2 -
\sum_{\alpha=1}^{N_p} \left( \i \breve{w}(\bm{r}_{\alpha}) + \i \sigma_{p,\alpha} \breve{\Phi}(\bm{r}_{\alpha})
\right) \right] \, ,  
\end{equation}
where $N_p$ is the number of beads in a polymer of type $p$.

The foregoing steps put us in a position to derive field operators 
whose ensemble averages correspond to the PDFs. 
First, consider the field operator
\begin{equation}
\tilde{\rho}_{\mathrm{b},p} (\bm{r}) \equiv \lim_{J_p\rightarrow 0} n_p
\frac{\delta \ln Q_p[\i \breve{w} - \breve{J}_p,\i \breve{\Phi} ]}{ \delta
J_p(\bm{r})} = \i n_p \frac{\delta \ln Q_p[\i \breve{w},\i \breve{\Phi}
]}{\delta w(\bm{r})} \, ,
\end{equation}
so named ($\sim \rho$) because
$\langle \tilde{\rho}_{\mathrm{b},p}
(\bm{r}) \rangle_{\mathrm{F}} = \langle \hat{\rho}_{\mathrm{b},p}(\bm{r})
\rangle_{\mathrm{P}}$. 
[Incidentally, this ensemble average is easily
computed by exploiting the translation invariance of the model. Since
$\langle\hat{\rho}_{\mathrm{b},p}(\bm{r}) \rangle_{\mathrm{P}} =
\langle\hat{\rho}_{\mathrm{b},p}(\bm{r}+\bm{a}) \rangle_{\mathrm{P}}$ for any
$\bm{a}$, $\langle\hat{\rho}_{\mathrm{b},p}(\bm{r})
\rangle_{\mathrm{P}} = \int \d \bm{r} \langle
\hat{\rho}_{\mathrm{b},p}(\bm{r}) \rangle_{\mathrm{P}}/V = \langle
\int \d \bm{r} \hat{\rho}_{\mathrm{b},p}(\bm{r}) \rangle_{\mathrm{P}}/V =
n_p N_p/V$, where $V$ is system volume.
The last equality holds because $ \int \d \bm{r}
\hat{\rho}_{\mathrm{b},p}(\bm{r}) = n_p N_p$ holds identically.] It should be emphasized that
the correspondence between this field operator and real-space
bead density exists only at the level of their respective ensemble 
averages. Although individual spatial configurations of the real 
part~\cite{FredricksonRev2002}
of $\tilde{\rho}_{\mathrm{b},p} (\bm{r})$ that is non-negative 
may be highly suggestive
and qualitatively consistent with the rigorous conclusions from PDFs 
(Fig.~2),
strictly speaking one cannot interpret
$\tilde{\rho}_{\mathrm{b},p} (\bm{r})$ in terms of
the actual bead positions for any single field configuration $\lbrace
w(\bm{r}),\Phi(\bm{r}) \rbrace$. 

We can compute $Q_p [\i \breve{w} ,\i \breve{\Phi} ]$ and
$\tilde{\rho}_{\mathrm{b},p} (\bm{r})$ for a given field configuration by
using so-called forward- and backward
chain propagators $q_{\mathrm{F},p}(\bm{r},\alpha)$ and
$q_{\mathrm{B},p}(\bm{r},\alpha)$, constructed iteratively using
the Chapman-Kolmogorov equations
\begin{align}
q_{\mathrm{F},p}(\bm{r},\alpha+1) &= \e^{-\i \breve{w}(\bm{r}) - \i
\sigma_{p,\alpha+1} \breve{\Phi}(\bm{r})} 
\left ( \frac {3}{2\pi b^2} \right)^{3/2}
\int \d \bm{r}' 
\e^{-3(\bm{r}-\bm{r}')^2/2 b^2} q_{\mathrm{F},p}(\bm{r}',\alpha) \, , \\
q_{\mathrm{B},p}(\bm{r},\alpha-1) &= \e^{-\i \breve{w}(\bm{r}) - \i
\sigma_{p,\alpha-1} \breve{\Phi}(\bm{r})} 
\left ( \frac {3}{2\pi b^2} \right)^{3/2}
\int \d \bm{r}' 
\e^{-3(\bm{r}-\bm{r}')^2/2 b^2} q_{\mathrm{B},p}(\bm{r}',\alpha) \, , 
\end{align}
while starting from 
$q_{\mathrm{F},p}(\bm{r},1) = \exp\left[ -\i \breve{w}(\bm{r}) -
\i \sigma_{p,1} \breve{\Phi}(\bm{r}) \right]$ and
$q_{\mathrm{B},p}(\bm{r},N_p) = \exp\left[ -\i \breve{w}(\bm{r}) - \i
\sigma_{p,N_p} \breve{\Phi}(\bm{r}) \right]$. With $q_{\mathrm{F},p}$ and
$q_{\mathrm{B},p}$ in place, we arrive at
\begin{align}
Q_p [\i \breve{w} ,\i \breve{\Phi} ]=\frac {1}{V}\int \d \bm{r} q_{\mathrm{F},p}
(\bm{r},N_p) \quad \mbox{and} \quad \tilde{\rho}_{\mathrm{b},p} (\bm{r}) =
\Gamma \star \frac{n_p}{V Q_p [\i \breve{w} ,\i \breve{\Phi} ]}
\sum_{\alpha=1}^{N_p} q_{\mathrm{F},p} (\bm{r},\alpha) q_{\mathrm{B},p}
(\bm{r},\alpha) \, \e^{ \i \breve{w}(\bm{r}) + \i \sigma_{p,\alpha}
\breve{\Phi}(\bm{r}) } \, .  
\end{align}

For inter-species PDFs, i.e., $G_{p,q}(|\bm{r}-\bm{r}'|)$ with $p \neq q$, 
Eq.~\eqref{eq:PDF_from_source} applied to Eq.~\eqref{eq:field_can_ensemble}
leads directly to
\begin{equation}
G_{p,q}(|\bm{r}-\bm{r}'|) = \langle \tilde{\rho}_{\mathrm{b},p} (\bm{r})
\tilde{\rho}_{\mathrm{b},q} (\bm{r}') \rangle_{\mathrm{F}} \; , \quad \quad p
\neq q \, .  \end{equation}
A direct application of Eq.~\eqref{eq:PDF_from_source} 
to obtain the intra-species PDF $G_{p,p}(|\bm{r}-\bm{r}'|)$ 
is also possible; but that procedure leads
to an expression containing a double functional derivative, viz.,
$\sim \delta^2 \ln Q_p/\delta w(\bm{r}) \delta w(\bm{r}') $, which 
is cumbersome to handle in numerical lattice simulations. 
We therefore obtain a simpler expression
by performing the field
redefinition $w(\bm{r}) \rightarrow w(\bm{r}) - \i J_p(\bm{r})$ instead
before taking
the second derivative. This alternate procedure
results in
\begin{equation}
G_{p,p}(|\bm{r}-\bm{r}'|) = \frac{\i}{v} \langle \tilde{\rho}_{\mathrm{b},p}
(\bm{r}) w (\bm{r}') \rangle_{\mathrm{F}} - \sum_{p \neq q} \langle
\tilde{\rho}_{\mathrm{b},p} (\bm{r}) \tilde{\rho}_{\mathrm{b},q} (\bm{r}')
\rangle_{\mathrm{F}} 
\; .
\end{equation}

In FTS, the continuum fields are approximated by discrete
field variables defined on a simple cubic lattice (mesh) with periodic boundary
conditions. Because of the complex nature of $H[w,\Phi]$, 
the Boltzmann factor $\exp(-H[w,\Phi])$ cannot be interpreted as a simple probability weight for a generic field configuration $\lbrace
w(\bm{r}),\Phi(\bm{r}) \rbrace$, which prohibits most standard Monte-Carlo techniques. This problem, known as the ``sign problem'', may be circumvented~\cite{FredricksonRev2002} by utilising a Complex-Langevin (CL) prescription \cite{ParisiWu1981,Parisi1983,Klauder1983,ChanHalpern1986}, where the fields are analytically continued into the complex plane. An artificial time coordinate $t$ is introduced and the fields evolve in CL-time according to the stochastic differential equations
\begin{equation} \label{eq:CL_evolution}
\frac{\partial \varphi(\bm{r},t)}{\partial t} = - \frac{ \delta
H}{\delta \varphi(\bm{r},t) }
+ \eta_{\varphi}(\bm{r},t)  \quad , \quad \varphi =w,\Phi \, . 
\end{equation}
Here, $\eta_{\varphi}$ represent real-valued Gaussian noise satisfying
$\langle \eta_{\varphi}(\bm{r},t)\rangle=0$ and
$\langle\eta_{\varphi}(\bm{r},t) \eta_{\varphi}(\bm{r}',t')\rangle = 
2 \delta(\bm{r}-\bm{r}') \delta(t-t')$. Thermal averages in the field picture can then be computed as asymptotic CL-time averages. In this work, we solve Eq.~\eqref{eq:CL_evolution} numerically using the first-order semi-implicit method of
\cite{LennonMohlerCenicerosGarcia-CerveraFredrickson2008}.

In computing PDFs in FTS, we can use knowledge of the translational and rotational invariance to make the computation more efficient. For instance, to calculate $ \langle
\tilde{\rho}_{\mathrm{b},p}(\bm{r}) \tilde{\rho}_{\mathrm{b},q}(\bm{r}')
\rangle_{\mathrm{F}}$, we can first calculate $\left\langle \int \d
\bm{a} \tilde{\rho}_{\mathrm{b},p}(\bm{r}+\bm{a})
\tilde{\rho}_{\mathrm{b},q}(\bm{r}'+\bm{a}) \right\rangle_{\mathrm{F}}/V $, 
which can be conveniently executed in Fourier space, with averaging 
over all possible directions
of $\bm{r}-\bm{r}'$. In this way, we obtain manifestly translationally and
rotationally invariant PDFs without spending computational time 
waiting for a droplet center of mass to explicitly visit all positions 
in the system or for a droplet to take on all possible spatial orientations. 
In the calculation of $G_{p,q}(|\bm{r}-\bm{r}'|)$
from lattice configurations, $|\bm{r}-\bm{r}'|$ is taken to be the 
shortest distance between positions $\bm{r}$ and $\bm{r}'$ with periodic 
boundary conditions taken into account.

The interplay of charge pattern and excluded volume
in the mixing/demixing of phase-separated polyampholyte species
is studied systematically for four
sequence pairs with $p$ $=$ sv28 ($-$SCD = $15.99$),
$q$ $=$ sv1, sv10, sv15, sv25 ($-$SCD = $0.41$, $2.10$, $4.35$, $12.77$), 
bulk monomer densities 
$\rho^0_{{\mathrm b},p}=\rho^0_{{\mathrm b},q}=0.25 b^{-3}$, and a 
moderately large $l_{\mathrm{B}} = 5b$ to ensure $T<$ critical temperature 
(see Sect. II above for rationale), each at 
excluded-volume strengths $v/b^3=0.0068$, $0.034$, $0.068$ and $0.102$. 
The latter three $v$ values are 5, 10 and 15 times the 
smallest $v/b^3=0.0068$, often used in FTS as
a relatively poor solvent condition 
\cite{LinMcCartyRauchDelaneyKosikFredricksonSheaHan2019,McCartyDelaneyDanielsenFredricksonShea2019,FredricksonPNAS}
favorable to LLPS~\cite{PerrySing2015}.
In this way, our analysis affords also a context for assessing 
the physicality of $v$ parameters
used commonly in FTS. As in recent works
\cite{LinMcCartyRauchDelaneyKosikFredricksonSheaHan2019,McCartyDelaneyDanielsenFredricksonShea2019},
we set the smearing length $a = b / \sqrt{6}$.

FTS in the present study is performed on $32^3$ and
$48^3$ lattices (meshes) with periodic boundary conditions and 
side-length $V^{1/3} = 13.88 b$ and $V^{1/3} = 24.0 b$, respectively. 
The Complex-Langevin (CL) evolution equations are integrated from random 
initial conditions using a step size $\Delta t = 0.001 b^3$ in CL time
for the $32^3$ mesh, and $\Delta t = 0.0005 b^3$ in CL time for the
$48^3$ mesh.
After an initial equilibration period of $40,000$ steps, the systems 
are sampled every 1,000 steps until a total of $\sim 1,000$ sample 
field configurations are obtained for each run. These field configurations
are used in the averages described above.
For each binary sequence mixture and excluded-volume strength $v$, 
$\sim 80$ and $\sim 40$ independent runs are performed, respectively, 
for the $32^3$ and $48^3$ systems.


\begin{figure*}[!ht]
\vskip 5mm
\begin{center}
\includegraphics[width=1.00\columnwidth]{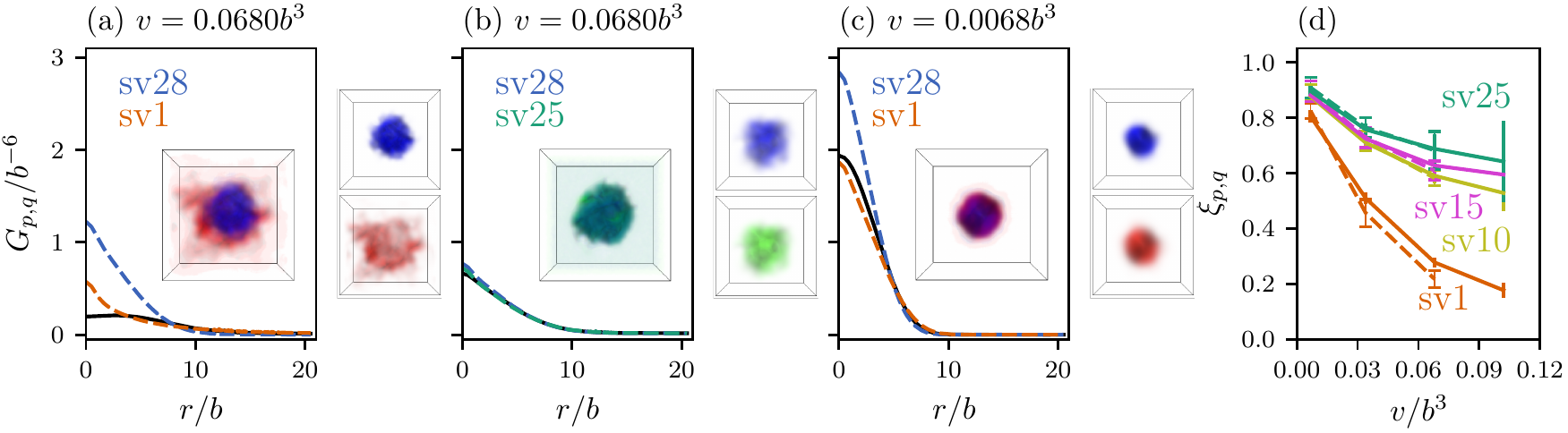}
\end{center}
\vskip -6mm 
\caption{FTS-computed PDFs and mixing parameter $\xi_{p,q}$ for binary
sv sequence mixtures. (a--c) Each $G_{p,p}$, $G_{q,q}$ (dashed, in color) and
$G_{p,q}$ (solid, black) for the indicated $v$ is
computed using a periodic $48^3$ mesh averaged over 30--40 independent 
runs (standard errors comparable to 
the plotting line width). Inset are illustrative snapshots of the 
real non-negative part of the density 
operators 
$\tilde{\rho}_{\rm{b},p}$ and $\tilde{\rho}_{\rm{b},q}$
depicted in different colors; the component species in the same snapshot
are shown separately on the side. 
(d) $\xi_{p,q}$ is computed using a periodic $32^3$ mesh
(averaged over 70--80 independent runs, solid lines) as well as the $48^3$ 
mesh (dashed lines) used for (a--c). Error bars represents standard 
errors of the mean.
}
\label{fig:FTS_correlations}
\end{figure*}


PDFs indicate that significant charge pattern mismatch and strong $v$
are both necessary for demixing.
Representative results are shown in Fig.~\ref{fig:FTS_correlations} 
(see Appendix and Supplemental Material for comprehensive results). 
The strongest demixing is observed
for sv28--sv1 with large charge pattern mismatch (SCDs differ by 15.58) 
at relatively high $v$ values; e.g., for $v=0.068 b^3$,
$G_{\mathrm{sv1},\mathrm{sv28}}(r)$ takes much lower values than
$G_{\mathrm{sv1},\mathrm{sv1}}(r)$ and $G_{\mathrm{sv28},\mathrm{sv28}}(r)$
as $r\to 0$ (Fig.\ref{fig:FTS_correlations}a), 
indicating that some of 
the $\mathrm{sv1}$ chains are expelled from the sv28-dense region. 
Even when a single droplet is formed, it harbors 
sub-regions where either $\mathrm{sv28}$ or $\mathrm{sv1}$ dominates
(snapshot in Fig.\ref{fig:FTS_correlations}a). However, when $v$ decreases 
to $0.0068 b^3$, all three $G$s for sv28--sv1 share similar profiles, 
implying that the common droplet is well mixed 
(Fig.\ref{fig:FTS_correlations}c).
In contrast, for sv28-sv25 with similar charge patterns
(SCDs differ by $3.22$), mixing in the phase-separated droplet remains
substantial even at higher $v$ (Fig.~\ref{fig:FTS_correlations}b). 
The general trend is summarized by the mixing parameter 
(Fig.~\ref{fig:FTS_correlations}d)
\begin{equation} \label{eq:xi_def}
\xi_{p,q} \equiv 
\frac{ 2\rho^0_{{\mathrm b},p}\rho^0_{{\mathrm b},q} G_{p,q}(0)}{(\rho^0_{{\mathrm b},q})^2 G_{p,p}(0) + 
(\rho^0_{{\mathrm b},p})^2 G_{q,q}(0) } 
\; ,
\end{equation}
which vanishes for two perfectly demixed species, because in
that case at least one of the factors in 
$ \hat{\rho}_{\mathrm{b},p}(\bm{r}) \hat{\rho}_{\mathrm{b},q}(\bm{r})$ would 
be zero for any $\bm{r}$, whereas $\xi_{p,q}=1$ when 
$ \hat{\rho}_{\mathrm{b},p}(\bm{r}) \propto \hat{\rho}_{\mathrm{b},q}(\bm{r})$,
i.e., when the species are perfectly mixed.


\begin{figure*}[!ht]
\centering
\includegraphics[width=.60\columnwidth]{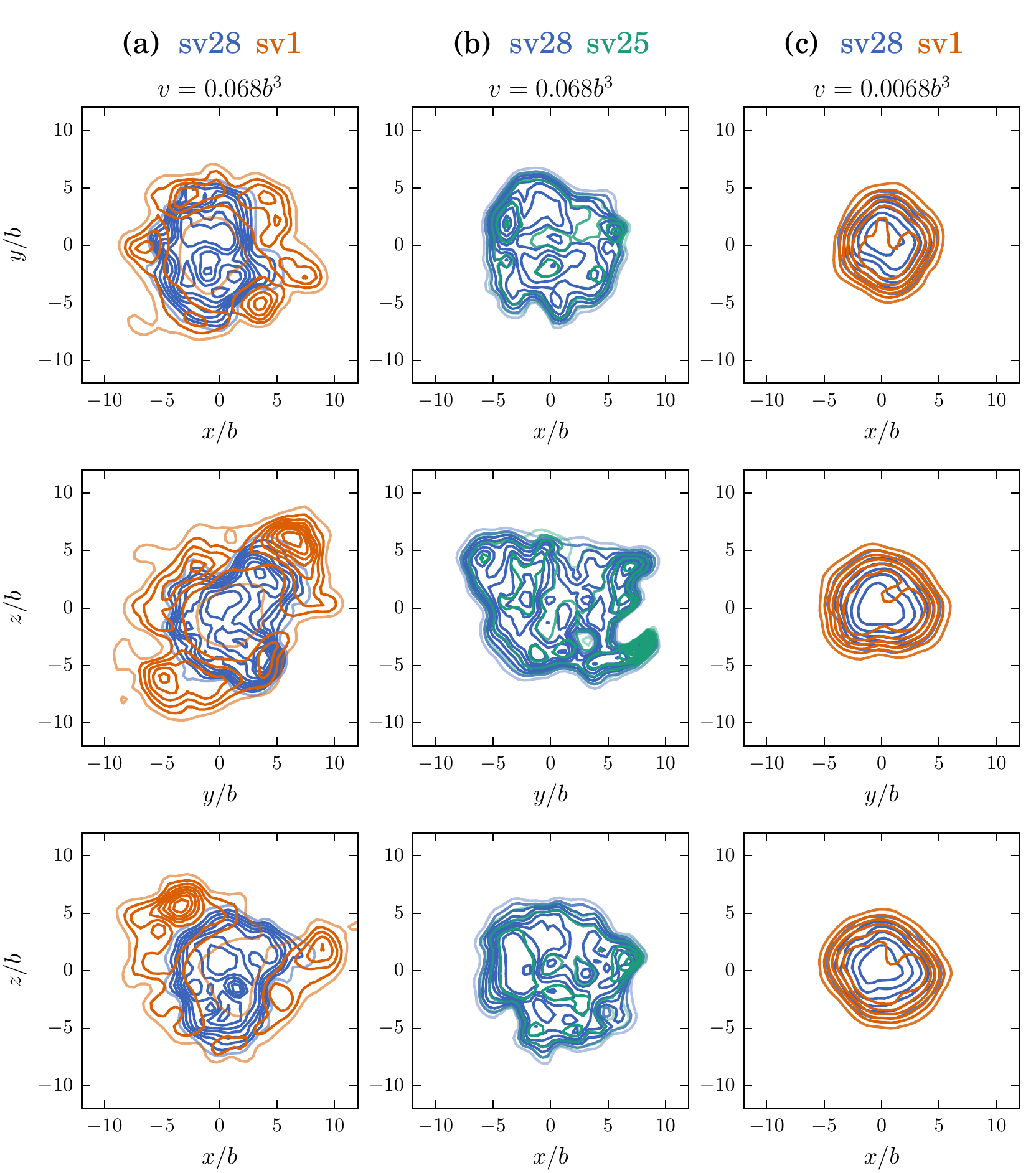}
\caption{Cross-sections of FTS droplets of binary sv sequence mixtures 
illustrating the interplaying roles of sequence charge pattern mismatch
and generic excluded volume in mixing/demixing of polyampholyte species.
Shown here are two-dimensional slides through the droplet center of mass
in the $x$--$y$ (top), $y$--$z$ (middle), and $x$--$z$ (bottom) planes
for the three FTS droplets depicted in Fig.~\ref{fig:FTS_correlations}a--c.
Density contours for the two sv sequence components $p,q$ in a given mixture
are color coded as indicated by the labels at the top of the (a)--(c) 
columns. The contours for species $p$ ($q$) are curves of constant bead 
density, where ``bead density'' here in a FTS snapshot means the 
real non-negative part of the density operator, viz., 
$\Re_+(\tilde{\rho}_{\mathrm{b},p/q} (\bm{r}))$
where
$\Re_+(u)\equiv [\Re(u)+{\rm sign}(\Re(u))]/2$ for any complex number $u$.
(Among all snapshots considered,
$\Re(\tilde{\rho}_{\mathrm{b},p} (\bm{r}))<-0.01b^{-3}$ occurs only for
$<2\%$ of the mesh points).
The contours are evenly spaced from 
$\Re(\tilde{\rho}_{\mathrm{b},p})$, 
$\Re(\tilde{\rho}_{\mathrm{b},q})$ $=$ $0$
[transparent] to 
$\Re_+(\tilde{\rho}_{\mathrm{b},p})$ $=$ 
$\max\{\Re_+(\tilde{\rho}_{\mathrm{b},p})\}$
($\Re_+(\tilde{\rho}_{\mathrm{b},q})$ $=$ 
$\max\{\Re_+(\tilde{\rho}_{\mathrm{b},q})\}$)
[opaque] where 
$\max\{\Re_+(\tilde{\rho}_{\mathrm{b},p})\}$
($\max\{\Re_+(\tilde{\rho}_{\mathrm{b},q})\}$)
is the maximum density of species $p$ ($q$) in a given snapshot.
}
\label{fig:FTS_snapshots_contours}
\end{figure*}


The dual requirements of a significant sequence charge pattern mismatch
and a substantial generic excluded volume for demixing of two polyampholyte
species in a condensed droplet are illustrated by the FTS snapshots for the 
sv28-sv1 pairs ($v/b^3=0.068$ and $0.0068$) 
and sv28-sv25 pairs ($v/b^3=0.068$) in Fig.~\ref{fig:FTS_correlations}a--c. 
Those snapshots present an overall view from the outside
of the droplet. Thus, part of their interior structure is obscured, albeit
this limitation is partly remedied by the translucent color scheme.
Further analyses to better understand the internal structures of these FTS 
snapshots are provided by the cross-sectional views in
Fig.~\ref{fig:FTS_snapshots_contours}. The contour plots in
Fig.~\ref{fig:FTS_snapshots_contours}a for the sv28-sv1 system with
a high generic excluded volume strength show clearly that there is indeed
a three-dimensional core with highly enriched sv28 population surrounded 
by a shell with enriched sv1 population. In contrast, the contour plots
for the sv28-sv25 system at the same excluded volume strength 
(Fig.~\ref{fig:FTS_snapshots_contours}b) and
the sv28-sv1 system at a low generic excluded volume strength
(Fig.~\ref{fig:FTS_snapshots_contours}c) 
indicate that the two polyampholytes species
are quite well mixed in the condensed droplets of these two systems. 
Nonetheless, the patterns of the contours reveals that even for these 
well-mixed systems, sv28 is 
still slightly more enriched in the core and the other sv sequence 
is slightly more enriched in a surrounding shell region.

\section{Explicit-chain coarse-grained molecular dynamics (MD) simulations}
While field theory affords deep physical insights, 
its ability to capture certain structure-related features pertinent 
to polyampholyte LLPS, such as the interplay between excluded volume
and Coulomb interactions, can be limited~\cite{DasAminLinChan}. 
To assess the robustness of the above FTS-predicted trend, we 
now turn to explicit-chain MD to
simulate binary mixtures of the same sv sequence pairs as with FTS, using 
an efficient protocol involving initial compression and subsequent 
expansion of a periodic simulation box for equilibrium
Langevin sampling~\cite{SHP2017,jeetainPLOS,DasAminLinChan}.
Each of our MD systems contains 500 chains equally divided 
between the two sv sequences (250 chains each).
The LJ parameter $\epsilon$ that governs excluded volume 
is set at $\epsilon=l_{\rm B}/a_0$
(corresponding to the ``with 1/3 LJ'' prescription in \cite{DasAminLinChan}),
$T^*\equiv(\beta\epsilon)^{-1}$ is reduced temperature, and a stiff
force constant $K_{\rm b}=75,000\epsilon/a_0^2$ for polymer bonds is 
employed as in \cite{SHP2017,DasAminLinChan}.
We compare results from using
van der Waals radius $r_0=a_0$ (as before \cite{DasAminLinChan}) and
$r_0=a_0/2$ to probe the effect of excluded volume.
Simulations are conducted at $T^*=0.6$ and $T^*=4.0$, which is,
respectively, below and above
the LLPS critical temperatures of all sv sequences
in Fig.~\ref{fig:sequences} in our MD systems, and at an 
intermediate $T^*$.

All MD simulations are performed using the GPU version of HOOMD-blue
simulation package \cite{ALT2008,GNALSMMG2015} as in
\cite{DasAminLinChan}.
For systems with excluded volume parameter $r_0=a_0$ (all systems
considered except in one case where we used $r_0=a_0/2$),
we initially randomly place all the polyampholyte chains inside a 
sufficiently large cubic simulation box of length $70a_0$.
The system is then energy minimized using the inbuilt FIRE algorithm to avoid
any steric contact for a period of $500 \tau$ with a timestep of $0.001 \tau$,
where $\tau \equiv \sqrt{ma^2/\epsilon}$ and $m$ is the mass of each bead
(representing a monomer, or residue).
Each system is first initiated at a higher temperature---at a 
high $T^*=4.0$---for a period of $ 5,000 \tau$.
The box is then compressed at $T^*=4.0$ for a period
of $5,000 \tau$ using isotropic linear scaling until we reach a sufficiently
higher density of $\sim 0.7 ma_0^{-3}$ which corresponds to a box size of 
$33a_0\times 33a_0\times 33a_0$. 
Next, we expand the simulation box length 
along one of the three Cartesian directions (labeled $z$)
8 times compared to its initial length to reach a final box length of
$264a_0$, hence the final dimensions of the box is
$33a_0\times 33a_0\times 264a_0$. 
For the system investigated for the effect of reduced excluded volume
with $r_0=a_0/2$, the initial compressed box size is 
$20a_0\times 20a_0\times 20a_0$, and the final box size is
$20a_0\times 20a_0\times 160a_0$. 
The box expansion procedure is conducted at a sufficiently
low temperature of $T^*=0.4$. After that, each system is equilibrated again at
the desired temperature for a period of $30,000 \tau$ using Langevin 
dynamics with a weak friction coefficient of $0.1m/\tau$ \cite{SHP2017}. 
Velocity-Verlet algorithm is used to propagate motion with periodic boundary
conditions for the simulation box. Production run is finally carried out 
for $100,000 \tau$ and molecular trajectories are saved every $10 \tau$ 
for subsequent analyses. 

For density distribution calculations, 
we first adjust the periodic simulation box in such a
way that its centre of mass is always at $z=0$. The simulation
box is then divided along the $z$-axis into 264 bins of size $a_0$ for
$r_0=a_0$ or 160 bins of size $a_0$ for $r_0=a_0/2$
to produce a total density profile as well as profiles for the two
individual polyampholyte species in the binary mixture. 
As for the $G_{p,q}(|\bm{r}-\bm{r}'|)$ in FTS,
in the calculation of the MD-simulated $G_{p,q}(|\bm{r}-\bm{r}'|)$
from configurations in the MD simulation box with periodic boundary 
conditions, $|\bm{r}-\bm{r}'|$ is taken to be the shortest distance of
the possible inter-bead distances determined in the presence
of periodic boundary conditions.


\begin{figure*}[!ht]
\begin{center}
\includegraphics[width=1.00\columnwidth]{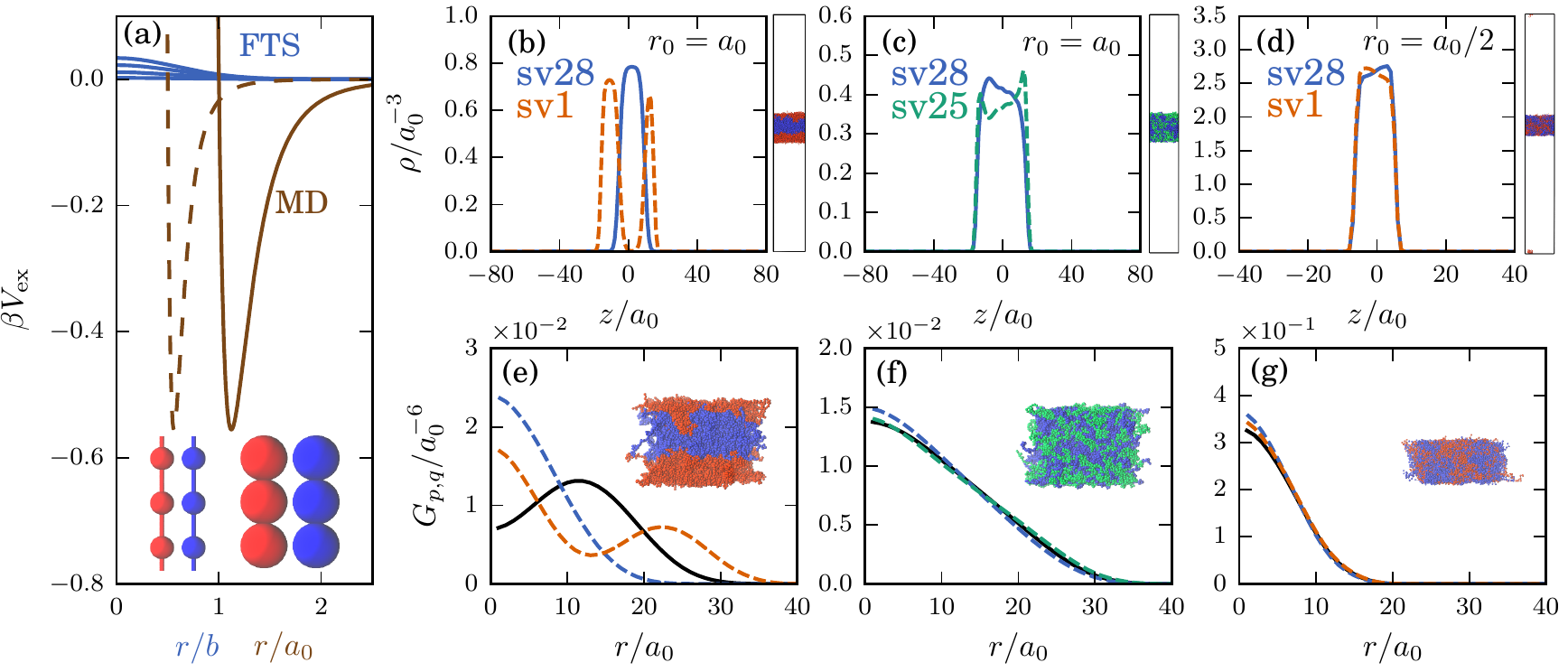}
\end{center}
\vskip -6mm 
\caption{MD-simulated LLPS of binary sv sequence mixtures.
(a) Excluded volume interactions
in FTS (blue) for $v/b^3$ $=$ $0.102$, $0.068$, $0.034$, 
and $0.0068$ (top to bottom, independent of $T^*$) 
and in MD (brown)
for $r_0=a_0$ (solid) and $r_0=a_0/2$ (dashed) at $T^*=0.6$ 
(insets show relative sizes of the LJ spheres).
(b)--(d) MD-simulated 
polyampholyte densities of binary mixtures, 
$\rho(z)$s for
different sv sequences are 
colored differently (as indicated) here and in
the snapshots (on the side) of the rectangular periodic simulation boxes
(wherein $z$ is the vertical coordinate), 
each harboring a condensed droplet.
(e)--(g) $G_{p,q}$ of the MD systems in (b)--(d), respectively,
(same line style as Fig.~\ref{fig:FTS_correlations}a--c).
Droplet snapshots (insets) are visualized~\cite{VMD} here with chains at 
periodic boundaries unwrapped.
}
\label{fig:explicit_chain_densities}
\end{figure*}


A substantive difference between common FTS and MD is in their 
treatment of polymer excluded volume, as illustrated in 
Fig.~\ref{fig:explicit_chain_densities}a for the present models,
wherein $\beta V_{\rm ex}(r)$ is the excluded-volume
interaction, given by $\beta {\hat H}_1$ in Table~\ref{tab1},
for a pair of beads centered at $\bm{r}_{p,i,\alpha}$ and 
$\bm{r}_{q,j,\gamma}$, with $r=|\bm{r}_{p,i,\alpha}-\bm{r}_{q,j,\gamma}|$.
For our FTS model as well as several recent FTS 
studies~\cite{McCartyDelaneyDanielsenFredricksonShea2019,LinMcCartyRauchDelaneyKosikFredricksonSheaHan2019,FredricksonPNAS},
\begin{equation}
\beta V_{\rm ex}(r)=
\frac{v}{2} \int \d \bm{r} \Gamma(\bm{r}-\bm{r}_{p,i,\alpha})
\Gamma(\bm{r}-\bm{r}_{q,j,\gamma})
=v \left( \frac{1}{4\pi a^2} \right)^{3/2} \exp\left( - \frac{r^2}{4a^2} \right) \quad \mbox{(FTS)} \, , 
\end{equation}
is a Gaussian, which allows~the~beads to overlap
completely ($r=0$), albeit with a reduced yet non-negligible or even
moderately high probability. In contrast, for MD, 
\begin{equation}
\beta V_{\rm ex}(r)=\frac{4}{3T^*} \left[ \left( \frac{r_0}{r} \right)^{12} - \left( \frac{r_0}{r} \right)^6 \right] \quad \mbox{(MD),}
\end{equation}
which entails a repulsive
wall at $\sim r_0$ that is all but impenetrable, let alone
an excluded-volume-violating complete overlap.
Note that if the 
$\beta V_{\rm ex}(r)$ for MD is shown for $T^*=0.2$ (as for FTS) instead
of $T^*=0.6$ in Fig.~\ref{fig:explicit_chain_densities}a, the contrast
would be even more overwhelming
between FTS and MD excluded-volume prescriptions.


\begin{figure*}[!ht]
\centering
\includegraphics[width=0.45\columnwidth]{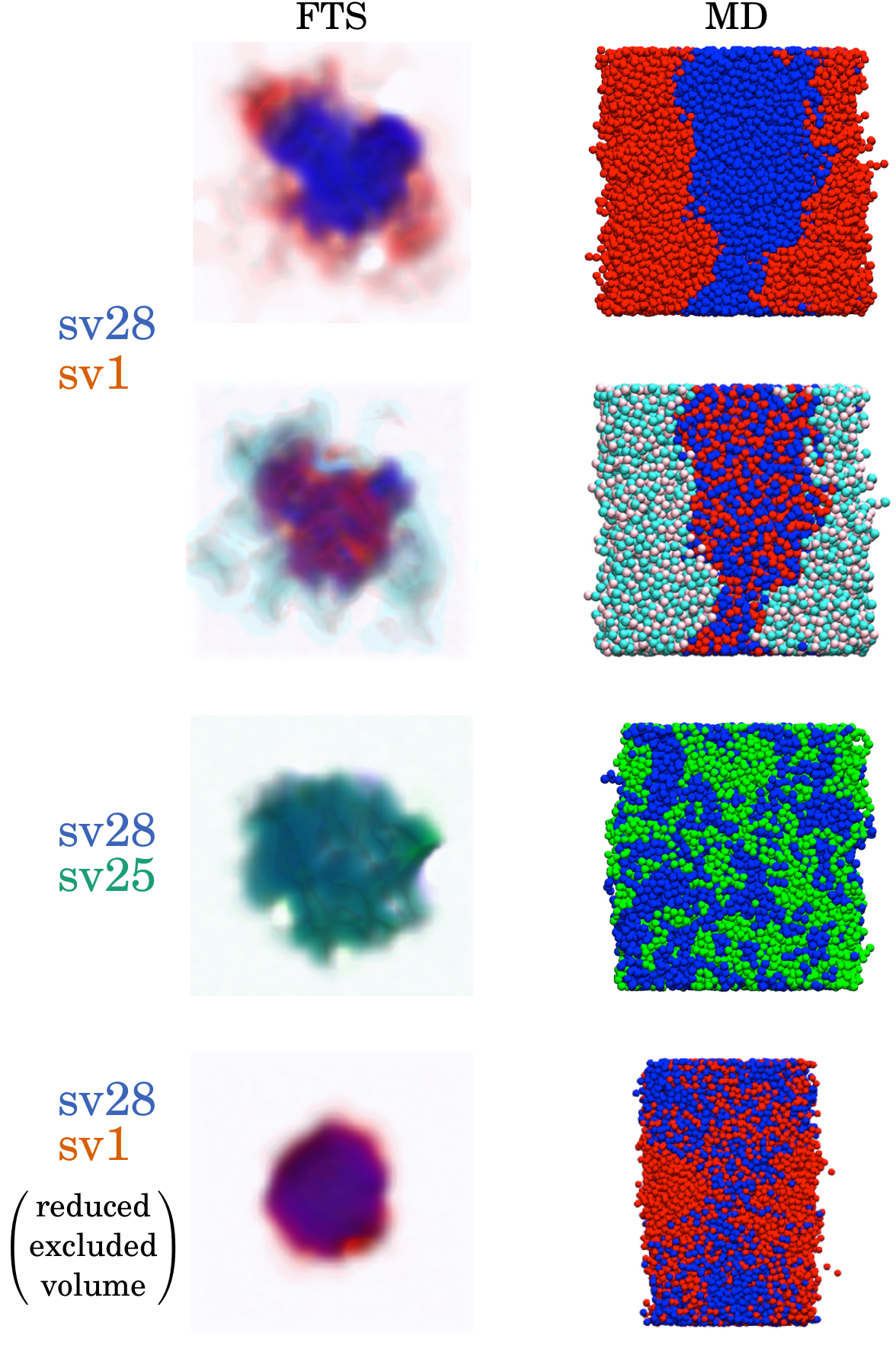}
\caption{Cross-sectional views of FTS and MD snapshots of binary mixtures 
of polyampholytes afford a consistent picture of sequence- and
excluded-volume-dependent droplet organization. 
(Left) FTS density distributions on one of the two-dimensional planes in 
Fig.~\ref{fig:FTS_snapshots_contours} through each droplet's center of mass.
(Right) Corresponding cut-out views of the MD droplets shown inside the 
periodic simulation boxes in Figs.~\ref{fig:explicit_chain_densities}b--d at one half of 
the box dimension extending perpendicularly into the page.
Two different representations are used to visualize the sv28-sv1 droplet 
with full excluded volume (top two rows; $v=0.068b^3$, $r_0=a_0$).
Upper row: sv1 and sv28 are depicted, respectively, in red and blue.
Lower row: The negatively and positively charged beads in sv28 are depicted,
respectively, in red and blue, whereas the corresponding beads in sv1 are
depicted in pink and cyan. The color code for
the sv28-sv25 mixture at full excluded volume (third row from top;
$v=0.068b^3$, $r_0=a_0$) and the sv28-sv1 mixture with reduced excluded 
volume (bottom row; $v=0.0068b^3$, $r_0=a_0/2$) follows that
in Figs.~\ref{fig:FTS_correlations} and \ref{fig:explicit_chain_densities}.
}
\label{fig:MD_FTS_snapshots}
\end{figure*}


Despite this and other 
differences between MD \cite{Sing2017,Whitmer2018,Sing2020} 
and field theory \cite{FredricksonPNAS,FredricksonJCP2019} that
preclude a direct comparison of MD and FTS excluded volume,
it is reassuring that MD and FTS 
predictions on sequence-pattern and excluded-volume dependent condensed-phase 
mixing/demixing share the same trend. 
Results for sv28--sv1 and sv28--sv25 are shown in 
Fig.~\ref{fig:explicit_chain_densities}b--g for $T^*=0.6$
to illustrate a perspective that is buttressed by additional 
MD results in the Appendix and Supplemental Material
for other sequence pairs, other $T^*$s,
their $\xi_{p,q}$ mixing parameters (Eq.~\ref{eq:xi_def}), and
$r_0=a_0/2$ sequences with proportionally reduced charged interactions.

Fig.~\ref{fig:explicit_chain_densities}b--d show the average
densities $\rho(z)$ along the long axis, $z$, of the 
simulation box. With full excluded volume and significant
charge pattern mismatch, sv28 and sv1 strongly demix in the condensed
phase (cf. blue and red
curves in Fig.~\ref{fig:explicit_chain_densities}b). In contrast,
without a significant charge pattern mismatch,
even with full excluded volume, sv28 and sv25 are quite well mixed
(blue and green curves largely overlap in 
Fig.~\ref{fig:explicit_chain_densities}c); and, with reduced
excluded volume, even sv28 and sv1 with significant
charge pattern mismatch are well mixed 
(Fig.~\ref{fig:explicit_chain_densities}d).

This trend is echoed by the PDFs
in Fig.~\ref{fig:explicit_chain_densities}e--g, each 
computed from 10,000 MD snapshots. For the well-mixed cases in 
Fig.~\ref{fig:explicit_chain_densities}f,g, the MD-computed self ($G_{p,p}$,
$G_{q,q}$) and cross ($G_{p,q}$) PDFs largely overlap,
similar to those in Fig.~\ref{fig:FTS_correlations}b,c for FTS.
For the sv28--sv1 pair with full excluded volume in MD, 
Fig.~\ref{fig:explicit_chain_densities}e 
shows that $G_{p,q}(r)$ is significantly
smaller than $G_{p,p}(r)$ and $G_{q,q}(r)$ for small $r$, as in
Fig.~\ref{fig:FTS_correlations}a for FTS. Here, the MD $G_{q,q}$ 
for sv1 exhibits a local maximum at $r\approx 23a_0$ corresponding
to the distance between two sv1 density peaks in 
Fig.~\ref{fig:explicit_chain_densities}b. This feature reflects 
the anisotropic nature of the rectangular simulation box adopted to 
facilitate efficient sampling~\cite{SHP2017}. Nonetheless,
the geometric arrangement of sv28 and sv1 in the MD system, 
as visualized by the snapshot in Fig.~\ref{fig:explicit_chain_densities}e, 
is consistent with that in Fig.~\ref{fig:FTS_correlations}a for FTS
in that an sv28-enriched core (blue) is surrounded by an sv1-enriched 
(red) periphery in both cases. The other MD snapshots
in Fig.~\ref{fig:explicit_chain_densities}f,g depict well-mixed droplets,
similar to the corresponding FTS snapshots in 
Fig.~\ref{fig:FTS_correlations}b,c.

The MD-simulated droplet snapshots at low temperature $T^*=0.6$
in Figs.~\ref{fig:explicit_chain_densities}b--g underscore that demixing of two 
polyampholyte species in a condensed droplet requires a significant 
mismatch in sequence charge pattern as well as a substantial
excluded volume repulsion. Because the beads (monomers) are represented 
in our MD drawings as opaque spheres, the bulk of those droplets
below the surface of the image presented cannot be
visualized. To better illustrate that the observed mixing/demixing trend
applies not only to the exterior of the presented image of those
droplets but persists in the parts underneath 
(as can be inferred by the behaviors of $G_{p,p}$, 
$G_{q,q}$, and $G_{p,q}$ in Figs.~\ref{fig:explicit_chain_densities}e--g), we prepare 
cut-out images of those droplets to reveal the spatial organization in
their ``core'' regions (Fig.~\ref{fig:MD_FTS_snapshots}). The spatial
configurations of the MD droplets and their general trend of behaviors
(Fig.~\ref{fig:MD_FTS_snapshots}, right column)
are very similar to those exhibited by cross-sectional views of 
FTS droplets (contour plots in Fig.~\ref{fig:FTS_snapshots_contours}
and density plots in Fig.~\ref{fig:MD_FTS_snapshots}, left column),
demonstrating once again the robustness of our observations.
By construction, MD provides much more spatial details than FTS in 
this regard. Of particular future interest is the manner in which
individual positively and negatively charged beads interact across
polyampholytes of different species. MD snapshots should be useful
for elucidating this issue. In contrast, although FTS
snapshots---with their cloudy appearances---may show a similar spatial 
organization of charge densities as that of MD, the field configurations 
do not translate into individual bead positions 
(Fig.~\ref{fig:MD_FTS_snapshots}, second row).


\section{Conclusion}
Excluded volume has been shown
to attenuate complex~\cite{PerrySing2015} and 
simple~\cite{McCartyDelaneyDanielsenFredricksonShea2019} coacervation
(i.e., excluded volume generally disfavors demixing of solute 
and solvent) but to promote demixing of molecular (solute) 
components when applied differentially to different molecular 
components in a condensate~\cite{harmonNJP}. Here, going beyond
these and other effects of excluded volume on the organization of 
condensed matter (e.g., nanogel~\cite{adro2017} and 
polymer-nanoparticle systems~\cite{sorichetti2018}),
FTS and MD both demonstrate a hitherto unrecognized stochastic molecular 
recognition principle, that a uniform excluded volume not discriminating
between polymer species can nonetheless promote condensed-phase demixing 
and that a certain threshold excluded volume is required for heteropolymers
with different sequence charge patterns to demix upon LLPS. 
Our MD results
show clearly that sequences such as sv28 and sv1 that are not obviously 
repulsive to each other can nevertheless demix in the condensed phase,
supporting RPA predictions that such demixing of different species of
overall neutral polyampholytes depends on charge pattern 
mismatch~\cite{LinNewJPhys}. In light of the present finding, this success
of RPA in~\cite{LinNewJPhys} may be attributed to the incompressibility 
constraint---which presupposes excluded volume---in its formulation.
Surprisingly, although the FTS excluded volume repulsion we consider 
is exceedingly weak---the highest $v$ only amounts to 
$\sim 0.03 k_{\mathrm{B}}T$ maximum
and thus can easily be overcome by
thermal fluctuations (Fig.~\ref{fig:explicit_chain_densities}a),
the demixing observed in FTS with this $v$ is similar to that
in MD with a much stronger, more realistic excluded volume.
While the theoretical basis of this reassuring agreement, e.g., its 
possible relationship with the treatment of chain entropy in FTS, 
remains to be ascertained, our observation that sv28 and sv1 do not demix
at a lower $v$ points to potential limitations of employing small $v$ 
values in FTS.

These basic principles offer new
physical insights into subcompartmentalization of membraneless 
organelles, in terms of not only the sequence charge patterns of 
their constituent IDPs~\cite{LinNewJPhys}, but also of excluded volumes 
entailed by amino acid sidechains of various sizes, volume increases due to 
posttranslational modifications such as phosphorylations~\cite{julie2019},
presence of folded domains, and the solvation properties of the IDP linkers 
connecting these domains~\cite{Rosen12,harmonNJP}. Guided by this
conceptual framework, quantitative applications to real-life biomolecular 
condensates require further investigations to consider
sequences that are not necessarily overall charge neutral~\cite{Kings},
and to incorporate non-electrostatic driving forces for LLPS
such as $\pi$-related~\cite{Robert-Julie} and 
hydrophobic~\cite{panag2020,zheng2020} interactions.
Much awaits to be discovered.
\\

\centerline{\bf ACKNOWLEGMENTS}

We thank Yi-Hsuan Lin for insightful discussions, and gratefully
acknowledge support by Canadian Institutes of Health Research grant
NJT-155930, Natural Sciences and Engineering Research Council 
of Canada Discovery grant RGPIN-2018-04351, and computational
resources from Compute/Calcul Canada.

T.P. and J.W. contributed equally to this work.

\vfill\eject


\renewcommand{\theequation}{{\rm A}\arabic{equation}}
\setcounter{equation}{0}

\begin{center}
{\bf APPENDIX: COMPREHENSIVE FTS AND MD RESULTS,
PAIR\\ CORRELATION FUNCTIONS AND SECOND VIRIAL COEFFICIENTS}
\end{center}

$\null$
\vskip -4mm

In this Appendix, figures with number labels preceded by ``S'' 
refer to the figures in Supplemental Material.

\centerline{\bf A. Comprehensive FTS results}

$\null$
\vskip -4mm

Figs.~S1 
and S2 
show PDFs of all of the sv sequence pairs considered in the present work.
They are computed using, respectively, the 
$32^3$ and $48^3$ meshes under
various excluded volume strengths $v$. Results are available for
the highest $v/b^3=0.102$ we simulated for the $32^3$ mesh but not for
the $48^3$ mesh because equilibration is problematic for the larger mesh
at strong excluded volume. 
At the low temperature ($l_{\rm B}=5b$, $T^*=0.2$) at which these simulations are
conducted, a hallmark for the existence of a condensed droplet is
the decay of the $G_{p,p}$, $G_{q,q}$, and $G_{p,q}$ functions to
$\approx 0$ at $r\approx 10b$; and a significant demixing of the populations
of the two sequence species
is signaled by a substantially lower $G_{p,q}(r)$ ($p\neq q$), for small
$r\approx 0$, than both $G_{p,p}(r)$ and $G_{q,q}(r)$ in the same range
of $r$. 
The trends exhibited by the two sets of results in 
Figs.~S1 
and S2 
are consistent. They
indicate robustly that both a significant difference in sequence charge pattern
of the two polyampholyte species (difference decreases from the sv28-sv1 
to the sv28-sv25 pair) and a substantial excluded volume (relatively large
$v$ values) are required for appreciable demixing. This observation
corroborates the trend illustrated by the sv28-sv1 and sv28-sv25 examples 
and the $\xi_{p,q}$ measure presented in Fig.~\ref{fig:FTS_correlations}.
As a control, and not surprisingly, when FTS is conducted at a much
higher temperature of $T^*=20$ ($l_{\rm B}=0.05b$) 
in Fig.~S3 
, there is little
sequence dependence---as seen by the very similar behaviors of all 
$G_{p,p}(r)$, $G_{q,q}(r)$, and $G_{p,q}(r)$ among the sequence pairs
considered---and there is no droplet formation. Instead of converging
to zero at large $r$ as in Figs.~S1
and S2,
here all $G(r)$s converge to a finite (nonzero) 
value of 
$\langle \hat{\rho}_{\mathrm{b},p}\rangle_{\mathrm{P}}
\langle \hat{\rho}_{\mathrm{b},q}\rangle_{\mathrm{P}}\approx 0.05 b^{-6}$ 
at large $r$ 
in Fig.~S3 
for $p\neq q$ as well as $p=q$,
signalling a total lack of correlation between distant beads.
\\

\centerline{\bf B. Comprehensive MD results}

$\null$
\vskip -4mm

Fig.~S4 
shows the density profiles
of six sv sequence pairs (the same sv pairs analyzed using RPA 
in Ref.~\cite{LinNewJPhys}). At a sufficiently low temperature of
$T^*=0.6$, LLPS is observed for all $r_0=a_0$ systems simulated here, in that
a droplet, manifested as a density plateau, is observed (left column
of Fig.~S4
). At this
low temperature, demixing of the two species in the binary mixture
is clearly observed for sv28-sv1 and sv28-sv10, and nearly complete
mixing is observed for sv28-sv24 and sv28-sv25. Intermediate
behaviors that may be characterized as partial demixing---with sv28
slightly enriched in the middle and the other sequence species slightly
enriched on the two sides---are observed for sv28-sv15 and sv28-sv20.
The trend is also seen at intermediate temperatures ($T^*=1.4$--$2.3$).
However, in some of these cases, one of the polyampholytes either does
not (e.g.~sv1) or barely (e.g.~sv15) phase separate, as indicated by
the long ``tails'' of their density profile outside the central region (middle
column of Fig.~S4
).
Not unexpectedly, at a high temperature of $T^*=4.0$, none of
the simulated systems phase separates and the two species are 
mixed homogeneously throughout the simulation box (right column of
Fig.~S4). 

These trends are summarized 
quantitatively in Fig.~S5 
using essentially the same
$\xi_{p,q}$ parameter 
defined in Eq.~\eqref{eq:xi_def}. Consistent with the FTS results
in Fig.~\ref{fig:FTS_correlations}, demixing of condensed-phase polyampholyte 
species increases with sequence charge pattern mismatch and increasing
excluded volume. Representative snapshots of our MD-simulated systems are
shown in Fig.~S6.
To highlight the impact of excluded volume, the $\xi_{p,q}$ parameter
for the sv28--sv1, $r_0=a_0/2$ system 
with reduced excluded volume (Figs.~\ref{fig:explicit_chain_densities}d,g)
is also shown in Fig.~S5 (red cross), exhibiting once again that when
$r_0=a_0/2$, sv28 and sv1 remain well mixed (do not demix) when a droplet is 
formed at low temperature (Fig.~S7, top),
these sequences' significant difference in charge pattern notwithstanding,
as has been shown by the pair distribution functions 
$G_{p,q}(r)$ in Fig.~\ref{fig:explicit_chain_densities}g.

To explore the potential impact of a stronger electrostatic 
interactions at contact---because of the reduced excluded volume---on this
lack of demixing, we further simulate a control system in which the charge on
each bead of the polyampholyte chains is scaled by a factor of $1/\sqrt{2}$
such that the electrostatic interaction energy when two beads are in contact
in the $r_0=a_0/2$ system is the same as that in the original $r_0=a_0$
system. Simulation results of this control system show that aside from
minor differences, the two species---sv28 and sv1---remain 
well mixed in the phase-separated droplet
(Fig.~S7 , bottom-left). 
This result, together with the recognition that beads on polyampholytes with
$r_0=a_0/2$ can interdigitate because the bonds connecting the chains have
no excluded volume (Fig.~\ref{fig:explicit_chain_densities}a, inset) 
and therefore likely allow for more 
mixing of polyampholyte species, confirms once again that excluded volume, 
overall, is a prominent driving factor for demixing of polyampholyte 
species in the condensed phase.
\\

\centerline{\bf C. Pair distribution functions and second virial coefficients}

$\null$
\vskip -4mm

Virial expansion is a perturbative approach useful for studying
nonideal gas and dilute solution as it is a power series in
density (concentration)~\cite{Pathria}. The coefficient of the second
term in the expansion of mechanical or osmotic pressure, 
known as the second virial coefficient and often 
denoted as $B_{22}$ or $B_2$, may be expressed as
\begin{equation}
B_2 = 2\pi\int_0^\infty \d r \; r^2 \left( 1 - \e^{-\beta U_2(r)} \right )
= 2\pi\int_0^\infty \d r \; r^2 \left[ 1 - g_{\rm dil}(r) \right ]
\label{eq:B2_eq00}
\end{equation}
for an isotropic pairwise potential $U_2(r)$, and the second equality
follows when $g_{\rm dil}$ is the normalized radial distribution in the 
limit of infinite dilution because $g_{\rm dil}(r)=\exp[-\beta U_2(r)]$
[$g_{\rm dil}(r)\rightarrow 1$ as $r\rightarrow\infty$]~\cite{lenhoff1998}.
As such, $B_2$ is particularly useful for characterizing the interactions
between two otherwise isolated molecules~\cite{Dignon2018,alanamin2020};
but is insufficient for an accurate account at high densities or 
high solute concentrations because contributions involving third 
and higher orders in density are neglected.

In contrast, the pair distribution functions (PDFs) computed in this work
are exact (inasmuch as the finite-size model systems considered are
concerned). For this reason, and in this regard, the configurational
information contained in PDFs is superior to that of $B_2$. Our PDFs are 
nonperturbative, and therefore they provide an
accurate characterization of the mixing/demixing of polyampholytes
species in both the dilute and condensed phases. 
To further compare and contrast the 
PDFs [$G_{p,q}(r)$ defined in Eq.~\eqref{eq:cross_correlation_particle} ]
in the present formulation and $B_2$,
it is instructive to define an exact radial distribution function,
%
\begin{equation}
g_{p,q}(r) \equiv \frac{G_{p,q}(r)}{\rho^0_{{\rm b},p} \rho^0_{{\rm b},q}} \, 
\label{eq:B_eq0}
\end{equation}
for our FTS as well as MD systems. 
Unlike the aforementioned $g_{\rm dil}(r)$,  
here $g_{p,q}(r)$ is not restricted to the dilute phase. Using $g_{p,q}(r)$ in
place of $g_{\rm dil}(r)$ in Eq.~\ref{eq:B2_eq00},
we may construct a second virial coefficient-like quantity
\begin{equation}
{\widetilde{B}}^{(2)}_{p,q} \equiv \frac{1}{2} \int_0^{r_{\mathrm{max}}} \d r
\, V(r/L) \left[ 1-g_{p,q}(r) \right] \, , 
\label{eq:B_eq1}
\end{equation} 
where $L$ is the side length of the simulation box. 
As in a recent simulation study of biomolecular 
condensates~\cite{ChoiDarPappu2019}, $V(r/L)$ is used
to adapt the integration measure to the periodic boundary conditions of 
a cubic simulation box, where
\begin{eqnarray}
V(x) &=& \left\lbrace \begin{matrix}
4 \pi x^2           \, , &   0\leq x \leq 1/2 \, ,  \\
2 \pi x (3 - 4 x)   \, , & 1/2 < x \leq \sqrt{2}/2 \, ,  \\
2 x (3 \pi - 12 f_1(x) + f_2(x) ) \, , & \sqrt{2}/2 < x \leq \sqrt{3}/2 \, ,
\end{matrix} \right.            \\
f_1(x) &=& \tan^{-1} \sqrt{4 x^2 - 1} \, , \\
f_2(x) &=& 8 x \left\{ \tan^{-1} 
\left[ \frac{ 2 x (4 x^2-3) }{\sqrt{4 x^2 - 2} (4 x^2 + 1) } \right]\right\}\, .
\end{eqnarray}
The above equations are Eqs.~18 and 19 in 
Ref.~\cite{ChoiDarPappu2019} (note, however, that our $g_{p,q}(r)$ is 
different from their ${\tilde{g}}(r)$ because of different normalizations).

The $1-g_{p,q}(r)$ expressions (in the integrand of Eq.~\ref{eq:B_eq1})
for our FTS systems are provided 
in 
Fig.~S8 
and 
Fig.~S9. 
For these phase-separated systems, unlike the $1-g_{\rm dil}(r)$ in
Eq.~\ref{eq:B2_eq00},
$1-g_{p,q}(r)$ does not vanish at large $r$ because large $r$ invariably 
involves the dilute phase and hence these $g_{p,q}(r) \approx 0$, i.e.,
$1-g_{p,q}(r)]\approx 1$ for large $r$. Therefore, it is sensible to
restrict the integration in Eq.~\ref{eq:B_eq1} to the condensed phase,
which may be implemented approximately by introducing an upper limit, 
$r_{\rm max}$, on the integration.

For $r_{\mathrm{max}} \leq L/2$, the volume integral reduces to
the simple form
\begin{equation}
{\widetilde{B}}^{(2)}_{p,q} \equiv 2 \pi \int_0^{r_{\mathrm{max}}} \d r \, r^2
\left[ 1-g_{p,q}(r) \right] \, , \quad r_{\mathrm{max}} \leq L/2 \, .
\label{eq:B_eq2}
\end{equation}
For the present MD systems, the final simulation boxes are not cubic, and
the dimensions of the condensed phase is approximately $L^3$ where
$L$ is the length of the shorter side of the simulation box
($L=33a_0$ for $r_0=a_0$ systems, $L=20a_0$ for $r_0=a_0/2$ systems;
see Fig.~\ref{fig:explicit_chain_densities}b--g and discussion above). 
For this reason, 
$r_{\mathrm{max}} \leq L/2$ should be chosen for the MD systems.
More generally, $r_{\rm max}$ may either be chosen as a pre-selected distance
reflecting the size of the condensed droplet, or as the solution to
the equation~\cite{ChoiDarPappu2019}
\begin{equation}
g_{p,q}(r_{\mathrm{max}}) = 1 \, .
\label{eq:B_eq3}
\end{equation}

We have computed ${\widetilde{B}}^{(2)}_{p,q}$ for our phase-separated
FTS systems using different pre-selected $r_{\rm max}$
as well as $r_{\rm max}$s satisfying Eq.~\ref{eq:B_eq3}, and found that
${\widetilde{B}}^{(2)}_{p,q}$ is quite insensitive to
reasonable variation in the choice of $r_{\rm max}$ as long as the choice
captures approximately the
size of the condensed droplet. Examples in 
Fig.~S10 
show that ${\widetilde{B}}^{(2)}_{p,q}$ for $p\ne q$
(black symbols) deviates more from ${\widetilde{B}}^{(2)}_{p,p}$ and/or
${\widetilde{B}}^{(2)}_{q,q}$ (symbols in other colors) with increasing
sequence charge pattern mismatch and increasing excluded volume.
The trend is most apparent for sv28--sv1 (Fig.~S10 , top left)
as this system entails a large 
sequence charge pattern mismatch. The trend 
observed in 
Fig.~S10 
of increased deviation of 
${\widetilde{B}}^{(2)}_{p,q}$ for $p\ne q$ from those for $p=q$
with increasing excluded volume 
is echoed by the MD example in 
Fig.~S11 
as well.
These examples underscore the fact that the configurational information
afforded by the second virial coefficient-like quantity
${\widetilde{B}}^{(2)}_{p,q}$ is derived from $G_{p,q}(r)$ and therefore
${\widetilde{B}}^{(2)}_{p,q}$ and $G_{p,q}(r)$ carry similar messages; but 
because ${\widetilde{B}}^{(2)}_{p,q}$ involves an $r$-integration 
of $G_{p,q}(r)$, ${\widetilde{B}}^{(2)}_{p,q}$ averages out spatial details
and thus contains less structural information of the system.
As such, ${\widetilde{B}}^{(2)}_{p,q}$ is not as diagnostic as $\xi_{p,q}$
in probing mixing/demixing of polyampholyte components in the condensed phase
(cf. Fig.~\ref{fig:FTS_correlations}d and Fig.~S5).
For that matter, as an integrated quantity, the second virial coefficient
$B_2$ itself (Eq.~\ref{eq:B2_eq00}) also provides less configurational 
information than $g_{\rm dil}(r)$.

$\null$\\
$\null$\\

\bibliography{bib_HSC}

\vfill\eject

\renewcommand{\d}{\mathrm d}
\renewcommand{\i}{\mathrm i}
\renewcommand{\thefigure}{S\arabic{figure}}
\renewcommand{\thetable}{S\arabic{table}}
\renewcommand{\theequation}{{\rm S}\arabic{equation}}
\setcounter{figure}{0}
%




\begin{center}
{\Huge\bf Supplemental Material}\\ 
$\null$\\
{\Large\bf Supplemental Figures}\\
{\large\it for}\\
{\Large\bf ``Subcompartmentalization of polyampholyte species in 
organelle-like condensates is promoted by charge pattern mismatch 
and strong excluded-volume interaction''}
\\

$\null$\\

{\large\bf
Tanmoy~Pal,$^{1\dagger}$ Jonas~Wess\'en,$^{1\dagger}$
Suman~Das,$^1$ and Hue Sun Chan$^{1*}$
}\\

{\large\it
$^1$Department of Biochemistry, University of Toronto\\ Toronto, 
Ontario M5S 1A8, Canada\\
}
$\null$\\
--------------------------------------------------------------------------------------------------------------------\\
$^\dagger$ T.P. and J.W. contributed equally to this work
\\
$^*$ To whom correspondence should be addressed.\\
Email: {\tt chan@arrhenius.med.utoronto.ca}
\end{center}

\vfill\eject


\begin{figure*}[ht]
\centering
\includegraphics{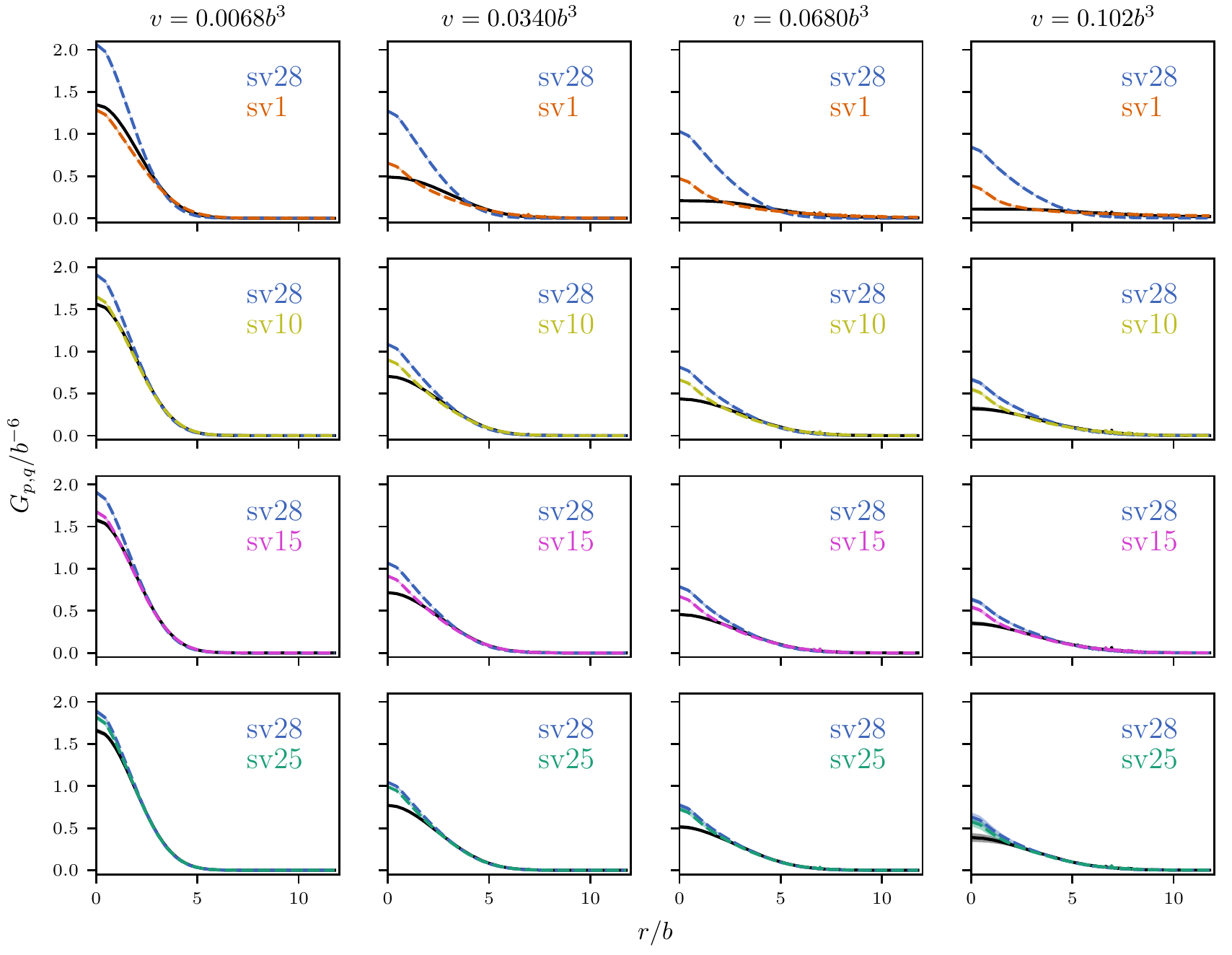}
\caption{PDFs of binary mixtures of sv sequences computed by FTS using a 
$32^3$ mesh at $l_{\rm B}=5b$ ($T^*=0.2$) and various $v$.
The plotting style follows that
of Fig.~2 
main text. Dashed blue curves: $G_{p,p}(r)$ for sv28
($-$SCD=$15.99$);
dashed color curves: $G_{q,q}(r)$ for (top to bottom) sv1, sv10, sv15, and
sv25 ($-$SCD=$0.41$, $2.10$, $4.35$, and $12.77$, respectively); 
solid black curves: $G_{p,q}(r)$. 
The shaded region around each curve represents 
standard error of the mean among the $\sim 80$ independent runs 
for each system, which is mostly 
smaller than the width of the curve. }
\label{fig:FTS_SI_32}
\end{figure*}

\vfill\eject



\begin{figure*}[!ht]
\centering
\includegraphics{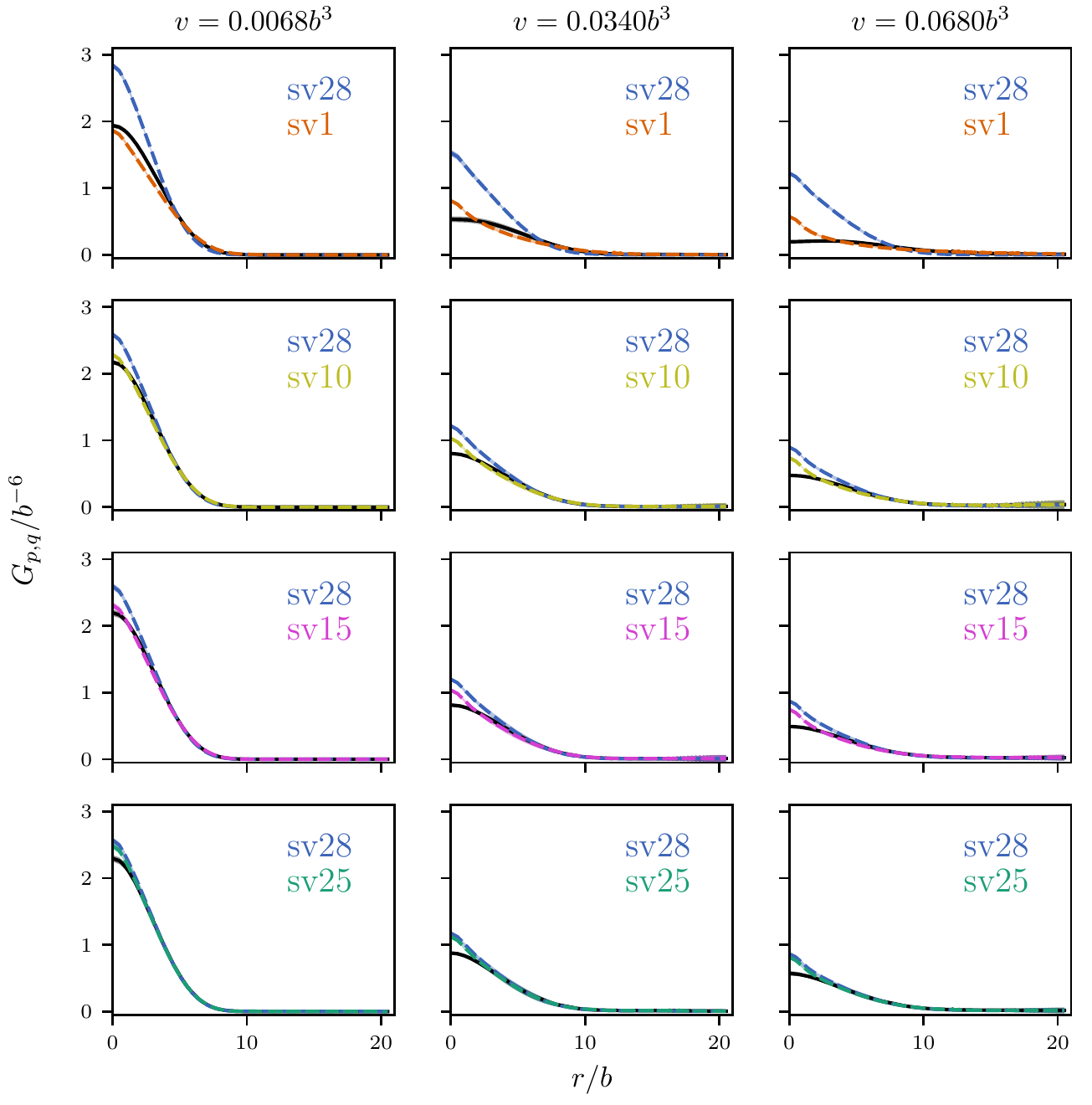}
\caption{PDFs of binary mixtures of sv sequences computed by FTS using a 
$48^3$ mesh at $l_{\rm B}=5b$ ($T^*=0.2$).
Results for each system are from $\sim 40$ independent runs.
The notation is otherwise the same as that of Fig.~\ref{fig:FTS_SI_32}.
}
\label{fig:FTS_SI_48}
\end{figure*}




\begin{figure*}[!ht]
\centering
\includegraphics{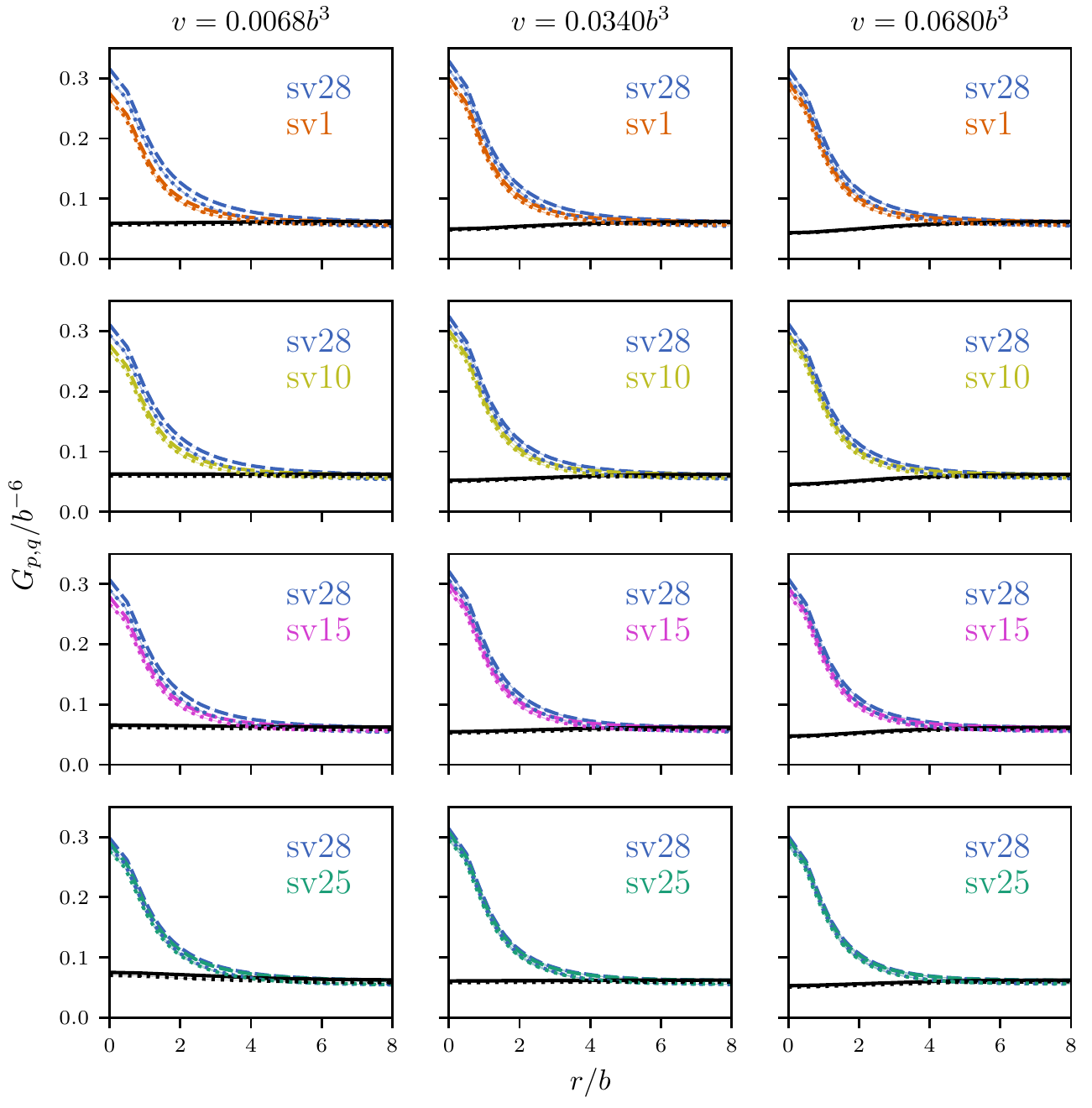}
\caption{PDFs of binary mixtures of sv sequences computed by FTS using a 
$32^3$ mesh or a $48^3$ mesh at $l_{\rm B}=0.05b$ ($T^*=20.0$).
Dashed (dotted) blue curves: $G_{p,p}(r)$ for sv28 from a
$48^3$ ($32^3$) mesh; dashed (dotted) color curves: $G_{q,q}(r)$ for 
(top to bottom) sv1, sv10, sv15, and sv25 from a $48^3$ ($32^3$) 
mesh; solid (dotted) black curves: corresponding $G_{p,q}(r)$
obtained using a $48^3$ ($32^3$) mesh.
At this high temperature, the behaviors of all systems are very
similar irrespective of the sequence charge patterns or excluded volume
interaction $v$ values considered.
The $r/b$ scale is enlarged
vis-\`a-vis Figs.~\ref{fig:FTS_SI_32} and
\ref{fig:FTS_SI_48} to make the differences between the plotted curves
here visible.
}
\label{fig:FTS_T20}
\end{figure*}

\vfill\eject



\begin{figure*}[!ht]
\centering
\includegraphics{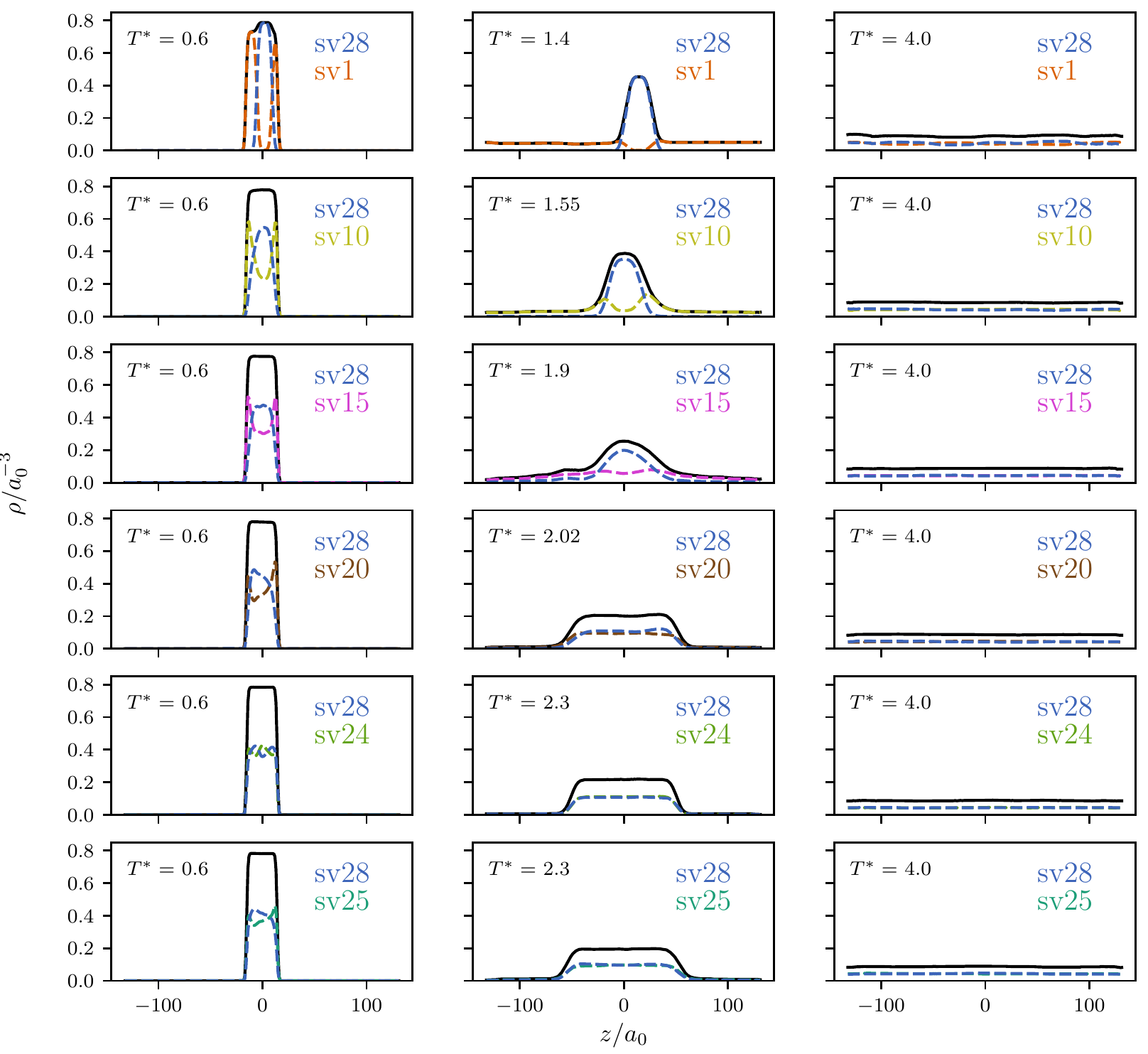}
\caption{MD-simulated average density of binary mixtures
of sv sequences along the $z$ (long) axis of the simulation box at
various temperatures for $r_0=a_0$. 
Solid curves: total bead density; color dashed curves: 
density of individual sv polyampholyte species. In addition to the
four sv pairs studied using FTS, MD results for the sv28-sv20 
($-$SCD = $15.99,7.37$)
and sv28-sv24 ($-$SCD = $15.99,17.00$) pairs are obtained to cover
the six sv pairs studied using RPA in
Y.-H. Lin, J. P Brady, J. D. Forman-Kay, and H. S. Chan, 
New J. Phys. {\bf 19}, 115003 (2017).
} 
\label{fig:explicit_chains_SI}
\end{figure*}

\vfill\eject



\begin{figure*}[!ht]
\centering
\includegraphics[width=13.8cm]{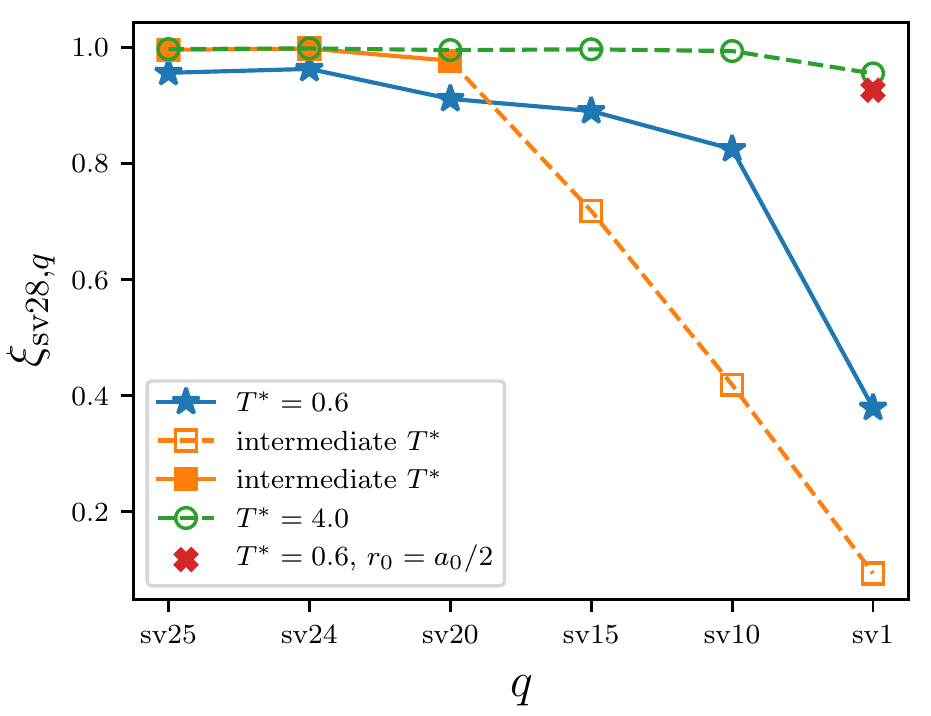}
\caption{$\xi_{p,q}$ values for the $p=$sv28, $r_0=a_0$ MD
systems in 
Fig.~\ref{fig:explicit_chains_SI}; with 
the $\xi_{{\mathrm{sv28}},{\mathrm{sv1}}}$
for $q=$sv1 and $r_0=a_0/2$
(Fig.~\ref{fig:explicit_chains_rescaled_charges_SI}, top-left) 
included for comparison (red cross). 
Since complete overlap of two beads ($r=0$) is all but impossible under the
$r^{-12}$ excluded-volume term in the MD potential, 
the $G_{p,q}(0)$ values used to
define $\xi_{p,q}$ in Eq.~16 of the main text are replaced by
the nonzero $G_{p,q}$ values at the smallest $r$ sampled (cf. Fig.~4e--f
of the main text) to compute the $\xi_{p,q}$ values plotted here.
Intermediate $T^*$ are those 
reported in the middle column of Fig.~\ref{fig:explicit_chains_SI}.
Systems with both polyampholyte species in the condensed phase are
depicted by fill symbols, those with one or both species
in the dilute phase
(not condensed) are represented by open symbols. The solid and
dashed connecting lines, respectively, through the filled and open symbols
serve merely as a guide for the eye.
Consistent with the FTS results in Fig.~2d of the main text, the
trend of $\xi_{p,q}$ values shown here indicates clearly that the 
demixing of two polyampholyte species 
in a condensed phase (at low $T^*$) increases with 
increasing mismatch of their sequence charge patterns, and decreases
with decreasing excluded volume (for sv1, the red cross indicates essentially
no demixing whereas the blue star indicates strong demixing in the
systems' respective combined condensed phases of sv28 and sv1 at $T^*=0.6$).
} 
\label{fig:MD_xi_SI}
\end{figure*}

\vfill\eject



\begin{figure*}[!ht]
\centering

\subfloat[sv28-sv1, $T^*=1.4$]{\includegraphics[width=17.2cm]{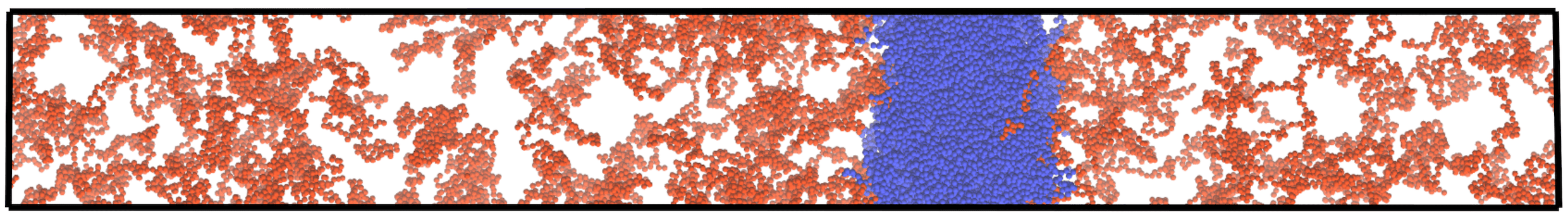}}

\subfloat[sv28-sv1, $T^*=4.0$]{\includegraphics[width=17.2cm]{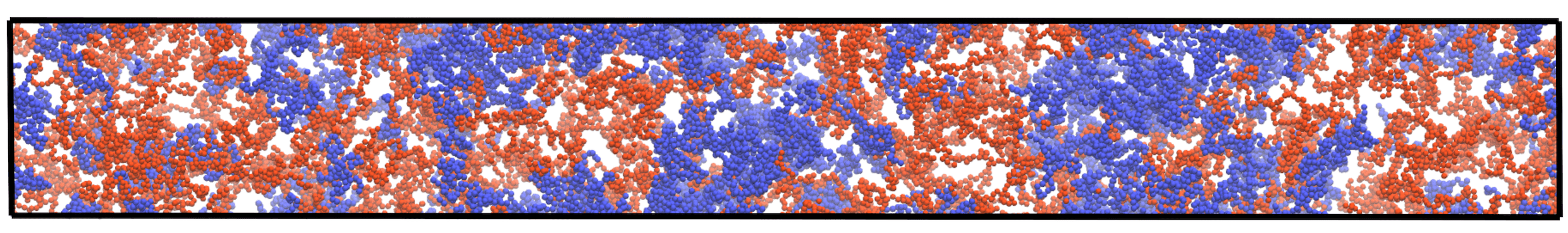}}

\subfloat[sv28-sv25, $T^*=2.3$]{\includegraphics[width=17.2cm]{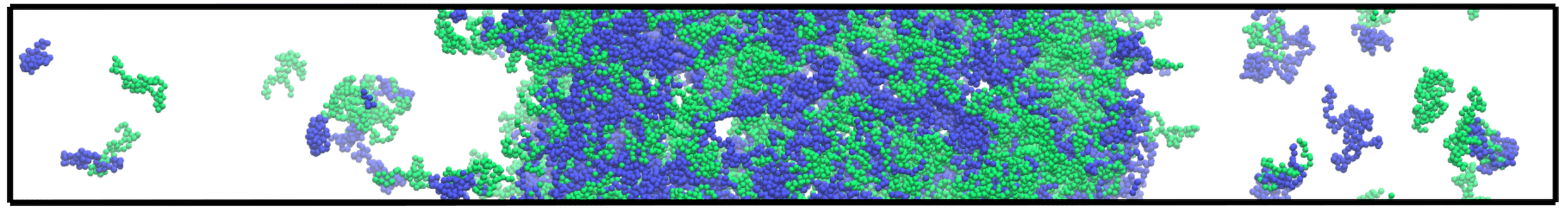}}

\subfloat[sv28-sv25, $T^*=4.0$]{\includegraphics[width=17.2cm]{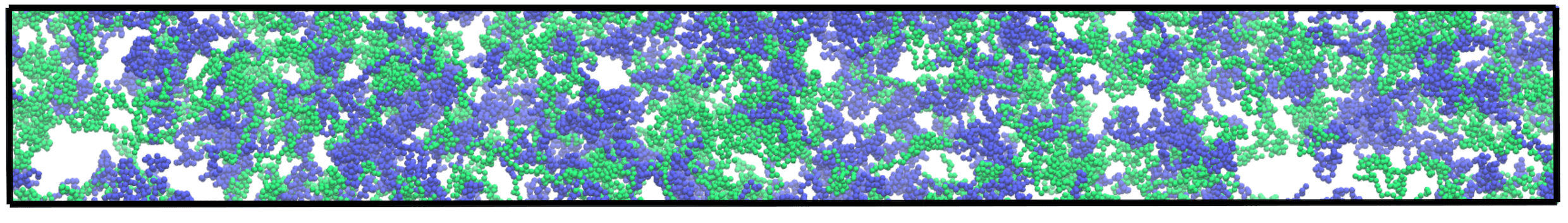}}

\caption{Simulation snapshots of binary mixtures of sv sequences 
at the reduced temperatures
indicated. Polyampholyte chains with charge sequences sv28, sv1, and sv25 
are depicted, respectively, in blue, red, and green.}

\label{fig:explicit_chains_SI_snapshots}
\end{figure*}

\vfill\eject



\begin{figure*}[!ht]
\centering
\includegraphics{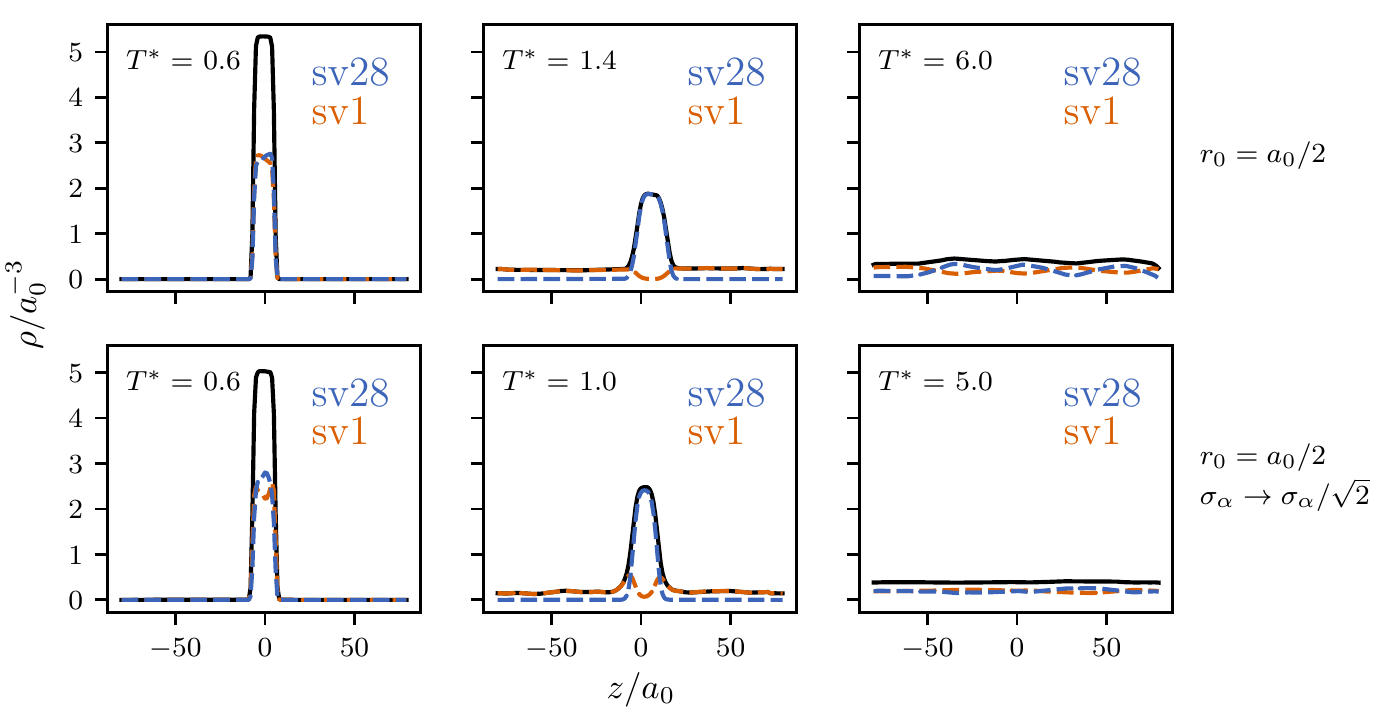}
\caption{MD-simulated average density of binary mixtures
of sv28 and sv1 along the $z$ (long) axis of the simulation box at
various temperatures for $r_0=a_0/2$. 
Solid curves: total bead density; color dashed curves: 
density of individual sv polyampholyte species; same line style 
as that in Fig.~\ref{fig:explicit_chains_SI}.
Top: Results from the $r_0=a_0/2$ system described in the main text.
Bottom: Results from a simulation wherein the electric charge on each
of the beads in the sv sequences is scaled by a factor of $1/\sqrt{2}$ 
such that the electrostatic interaction energies for two contacting beads
in this $r_0=a_0/2$ system is identical to that in
the original $r_0=a_0/2$ system with the original (unscaled) charges
on the beads.
} 
\label{fig:explicit_chains_rescaled_charges_SI}
\end{figure*}

\vfill\eject



\begin{figure*}[!ht]
\centering
\includegraphics[width=12.3cm]{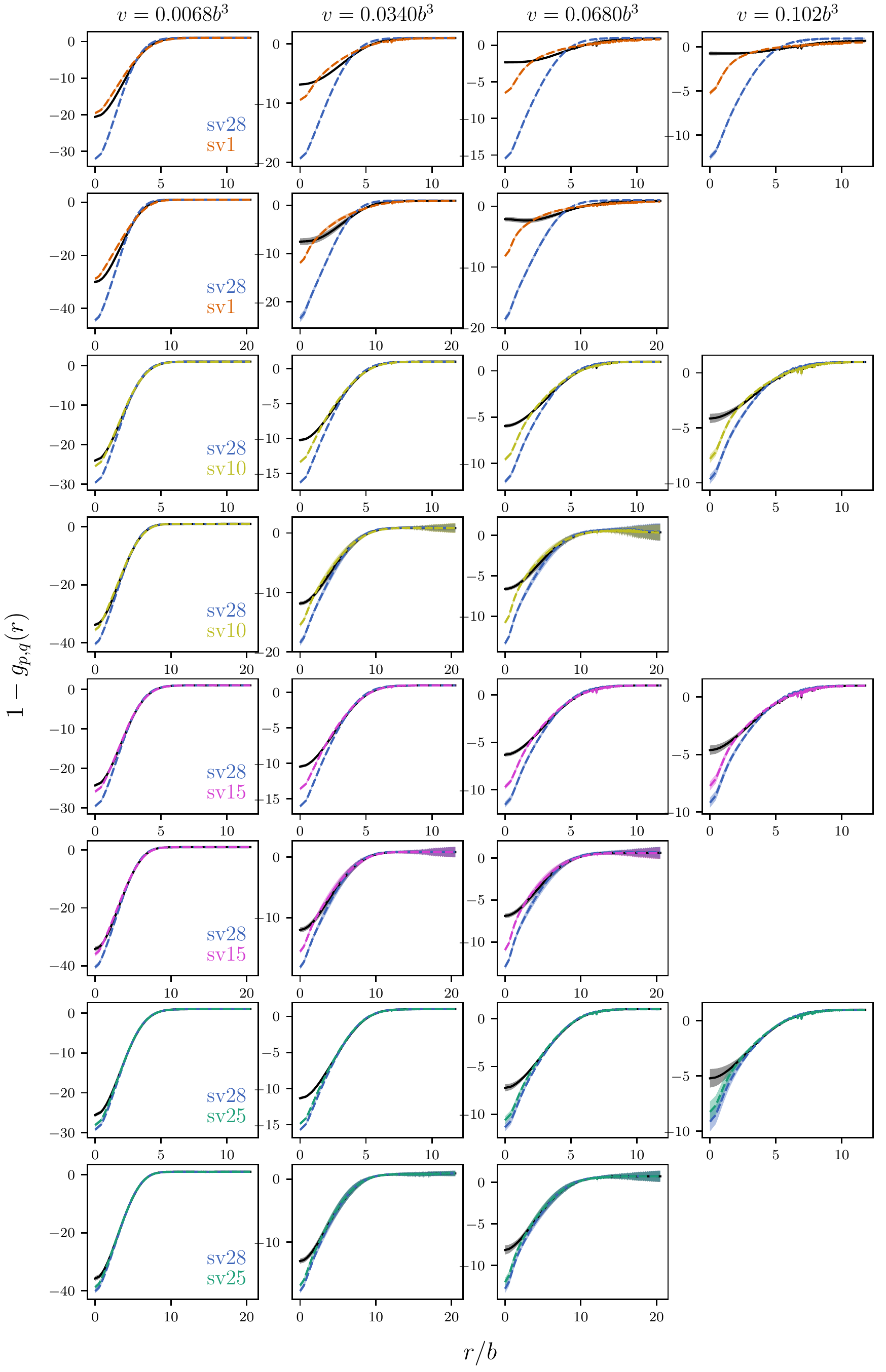}
\vskip -5mm
\caption{Radial distribution functions of FTS phase-separated systems
($l_{\rm B}=5b$, $T^*=0.2$).
Shown here is the function $1-g_{p,q}(r)$, where the normalized radial 
distribution $g_{p,q}(r)$ is defined by Eq.~A2 in the Appendix of the
main text
with the pair distribution functions (PDFs), $G_{p,q}(r)$,
given by Fig.~\ref{fig:FTS_SI_32} ($32^3$ mesh) and Fig.~\ref{fig:FTS_SI_48} 
($48^3$ mesh). The line styles for $p,q$ are the same as in
Figs.~\ref{fig:FTS_SI_32} and \ref{fig:FTS_SI_48}. Here,
for the same sv-sequence pairs, the
upper and the lower rows show $1-g_{p,q}(r)$ (which is the integrand
for ${\widetilde{B}}^{(2)}_{p,q}$ given by Eq.~A3 in the Appendix of the
main text)
for $32^3$ and
$48^3$ meshes, respectively. 
Shown results computed using the two different mesh sizes are
nearly identical.
The error bars follow those for the $G_{p,q}(r)$s, rescaled here in accordance
with Eq.~A2 in the Appendix of the main text.
} 
\label{fig:1-gr_FTS_1_SI}
\end{figure*}

\vfill\eject



\begin{figure*}[!ht]
\centering
\includegraphics[width=13cm]{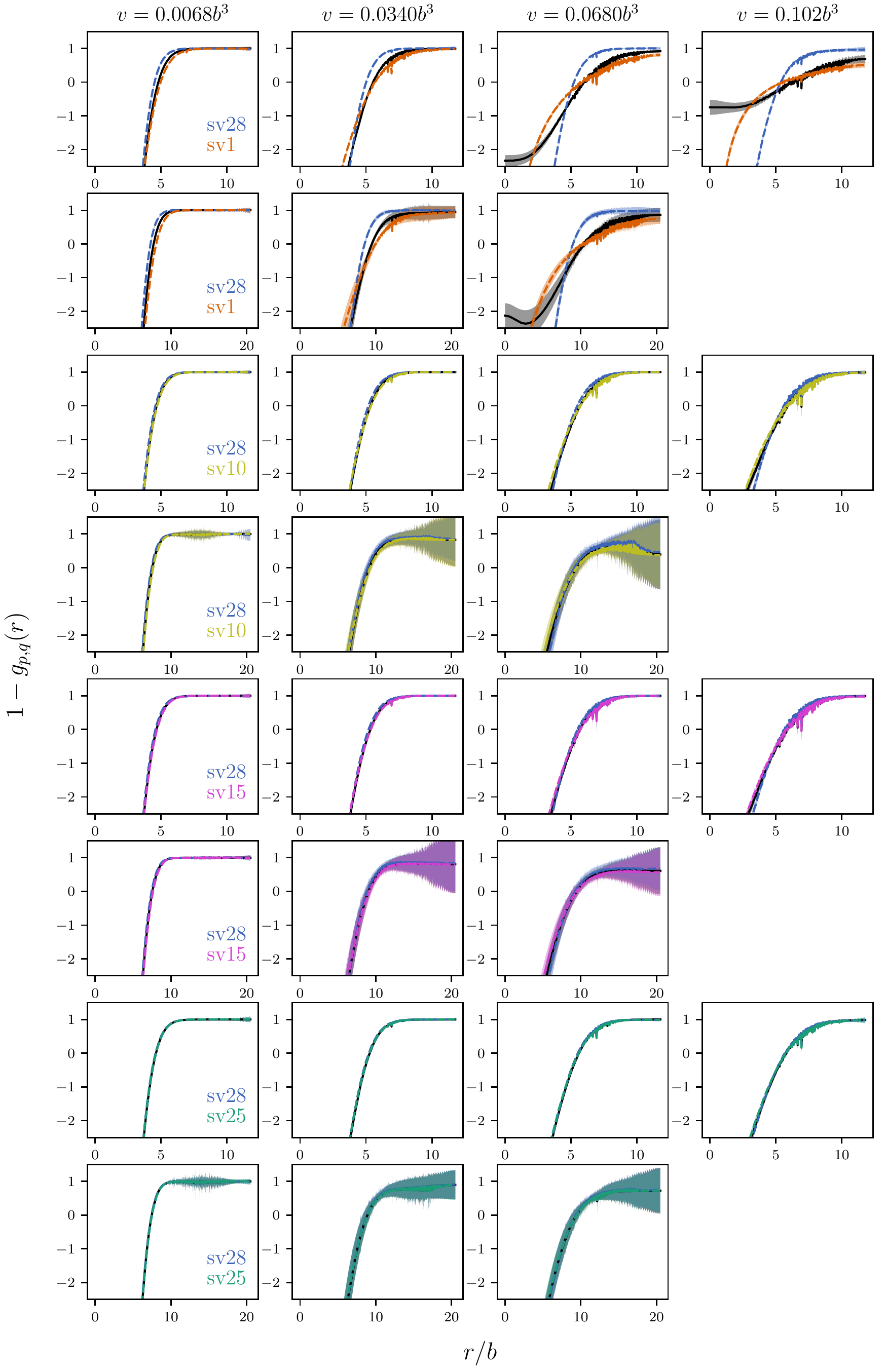}
\caption{Zoomed-in view of the radial distribution functions of FTS 
phase-separated systems. Same as 
Fig.~\ref{fig:1-gr_FTS_1_SI} but now with
a zoomed-in view around $1-g_{p,q}(r)=0$. In all cases, $1-g_{p,q}(r)\approx 1$
[i.e., $g_{p,q}(r)\approx 0$)] at large $r$.
} 
\label{fig:1-gr_FTS_2_SI}
\end{figure*}
\vfill\eject



\begin{figure*}[!ht]
\centering
\includegraphics[width=15cm]{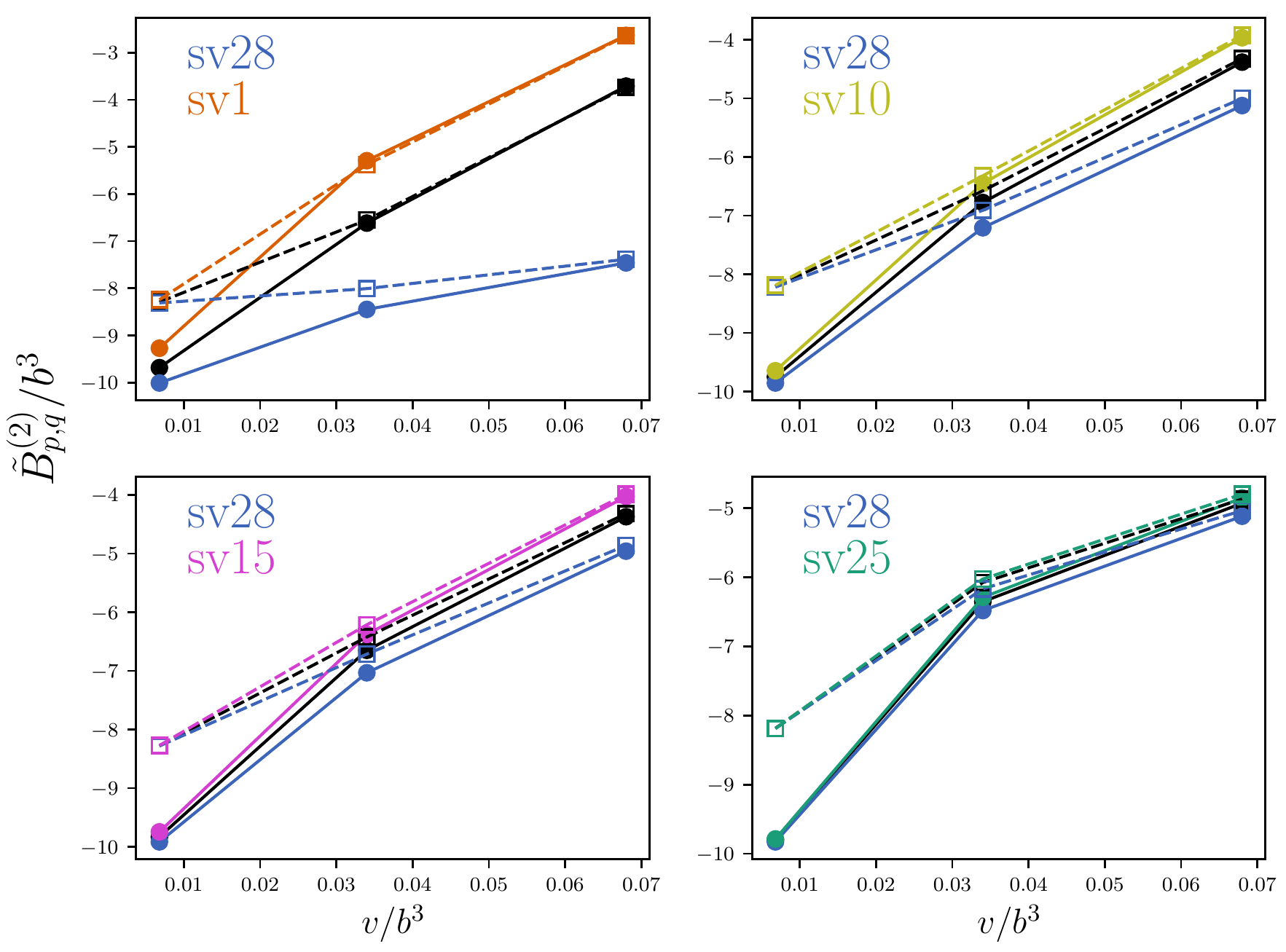}
\caption{The second virial coefficient-like quantity
${\widetilde{B}}^{(2)}_{p,q}$ for various phase-separated FTS systems 
($l_{\rm B}=5b$, $T^*=0.2$) each consisting
of two polyampholyte
species with different charge 
sequences (as indicated) computed using a $48^3$ mesh, as functions
of the excluded volume parameter $v$.
Filled circles represent ${\widetilde{B}}^{(2)}_{p,q}$ computed
by Eq.~A3 
using an $r_{\rm max}$ 
that equals to the average of the three $r_{\rm max}$ values satisfying
$g_{p,q}(r_{\rm max})=1$ (Eq.~A8)
for the $p,p$, $q,q$, and $p,q$ cases of the given pair of
sequences.
The open squares are ${\widetilde{B}}^{(2)}_{p,q}$ values computed
using $r_{\rm max}=10b$ throughout.
The solid and dashed lines connecting, respectively, the filled and open 
symbols are merely a guide for the eye.
} 
\label{fig:B_2_FTS_SI}
\end{figure*}
\vfill\eject


\begin{figure*}[!ht]
\centering
\includegraphics[width=17.2cm]{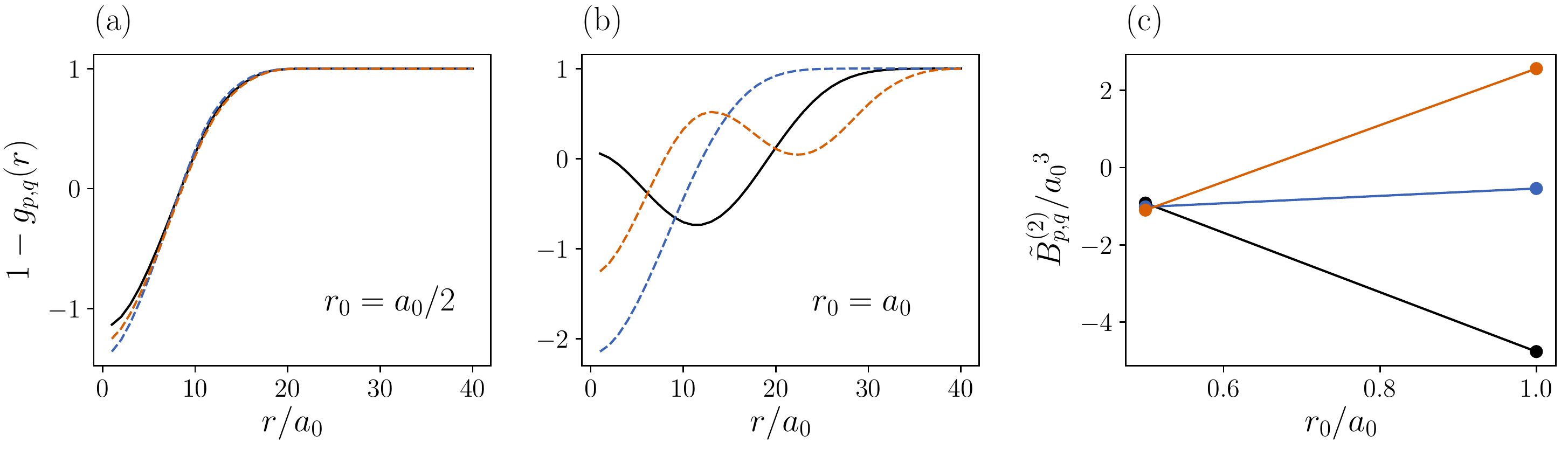}
\caption{Comparing MD-simulated ${\widetilde{B}}^{(2)}_{p,q}$ 
of phase-separated systems with different excluded volume strengths.
(a,b) The $1-g_{p,q}(r)$ functions for the sv28--sv1 MD systems 
are obtained, respectively, from 
the $G_{p,q}(r)$ functions in Fig.~4g ($r_0=a_0/2$) and Fig.~4e 
($r_0=a_0$) of the main text
(Eq.~A2). 
The line styles for $p,p$, $q,q$, and $p,q$ 
are the same as in the main text figure.
(c) The corresponding ${\widetilde{B}}^{(2)}_{p,q}$ values (same color code) 
are computed using $r_{\rm max}=10.0 a_0$ for $r_0=a_0/2$ and
$r_{\rm max}=16.5 a_0$ for $r_0=a_0$. These $r_{\rm max}$ values amount
to half of the length of the short sides of the systems' simulation boxes.
The solid lines connecting the $r_0=a_0/2$ and $r_0=a_0$ 
${\widetilde{B}}^{(2)}_{p,q}$s are merely a guide for the eye. 
} 
\label{fig:B_2_explicit_chain_SI}
\end{figure*}

\vfill\eject



\end{document}